\documentclass[11pt, letterpaper]{article} 

\usepackage[includeheadfoot,
            marginratio={1:1,2:3}, 
            width=440pt, 
            height=688pt,]{geometry}

\usepackage{amsmath}
\usepackage{amsfonts}
\usepackage{amssymb}
\usepackage{bbm,bm}
\usepackage{mathtools}
\usepackage{slashed}
\usepackage{mathalfa}

\usepackage{graphicx,caption,subfigure}
\usepackage{tikz} 
\usepackage{tikz-cd}
\usetikzlibrary{decorations.pathreplacing,calc,tikzmark}
\usepackage{booktabs}
\usepackage{adjustbox}
\usepackage[utf8]{inputenc}
\usepackage[splitrule]{footmisc} 
 
\usepackage{placeins}
\usepackage{empheq}
\usepackage{paralist}
\usepackage{cite}
\usepackage[normalem]{ulem}
\usepackage{soul}

\usepackage[colorlinks=false,urlbordercolor=red
]{hyperref}
\usepackage{xcolor}
\usepackage{tensor}
\usepackage{cancel}
\usepackage{orcidlink}

\newcommand{\nc}{\newcommand}
\nc{\lb}{\llbracket}
\nc{\rb}{\rrbracket}
\nc{\gl}{\llbracket}
\nc{\gr}{\rrbracket}
\nc{\del}{\partial}
\nc{\eq}[1]{\begin{equation}
                     \begin{split} #1 \end{split}
                     \end{equation}
}
\nc{\ov}{\overline}
\nc{\fa}{\hat}
\nc{\fb}{\MakeUppercase}
\nc{\fc}{\tilde}
\nc{\myhash}{\raisebox{\depth}{\#}}

\allowdisplaybreaks[2]
\numberwithin{equation}{section}
\DeclareUnicodeCharacter{2212}{-}

\begin{document}

\renewcommand*{\thefootnote}{\fnsymbol{footnote}}
\hypersetup{pdfborder={0 0 0}}

\vspace*{-1.5cm}
\begin{flushright}
  {\small
  MPP-2025-53\\
  LMU-ASC 10/25
  }
\end{flushright}

\vspace{1.0cm}
\begin{center}
  {\huge Emergence  of $F^4$-couplings in \\[0.3cm] Heterotic/Type IIA Dual
    String Theories} 
\vspace{0.4cm}

\end{center}

\vspace{0.25cm}
\begin{center}
{
  \Large Manuel Artime$^{1,2}$\footnote{e-mail: \href{mailto:artime@mpp.mpg.de}{artime@mpp.mpg.de}}\orcidlink{0009-0008-9385-2638}, Ralph Blumenhagen$^{1}$\footnote{e-mail: \href{mailto:blumenha@mpp.mpg.de}{blumenha@mpp.mpg.de} (corresponding author)}\orcidlink{0000-0001-7889-0603} and \\[0.15cm]
Antonia Paraskevopoulou$^{1,2}$\footnote{e-mail: \href{mailto:aparaske@mpp.mpg.de}{aparaske@mpp.mpg.de}}\orcidlink{0009-0001-3570-2518}
}
\end{center}

\vspace{0.0cm}
\begin{center} 
\emph{
$^{1}$ 
Max-Planck-Institut f\"ur Physik (Werner-Heisenberg-Institut), \\ 
Boltzmannstr.  8,  85748 Garching, Germany } 
\\[0.1cm] 
\vspace{0.25cm} 
\emph{$^{2}$ Fakult{\"a}t f{\"u}r Physik, Ludwig-Maximilians-Universit{\"a}t M\"unchen, \\ 
Theresienstr.~37, 80333 M\"unchen, Germany}\\[0.1cm]
\vspace{0.3cm}
\end{center} 
\vspace{0.5cm}

\begin{abstract}
The M-theoretic emergence proposal claims that in an isotropic decompactification limit to M-theory the full effective action is generated via quantum effects by integrating out only the light towers of states of the theory. In the BPS particle sector, these include transversally wrapped $M2$- and $M5$-branes possibly carrying Kaluza-Klein momentum. This implies that a longitudinally wrapped $M5$-brane, i.e. a wrapped $D4$-brane, is not to be included in emergence computations. In this work we collect explicit evidence supporting this point by examining an $F^4$ gauge coupling in six dimensions,
making use of the duality between heterotic string theory on $T^4$ and strongly coupled type IIA on K3. In this instance, the M-theoretic emergence proposal can be viewed as a tool for making predictions for the microscopic behavior of string theoretic amplitudes. 
\end{abstract}

\thispagestyle{empty}
\clearpage

\setcounter{tocdepth}{2}


\renewcommand*{\thefootnote}{\arabic{footnote}}
\setcounter{footnote}{0}
\hypersetup{pdfborder={0 0 1}}

\newpage
\section{Introduction}

The main goal of the swampland program (see e.g.\cite{Palti:2019pca,vanBeest:2021lhn, Grana:2021zvf,Agmon:2022thq} for reviews) is to determine which effective field theories can admit a consistent ultraviolet completion into quantum gravity theories. Despite providing insights to restrictions that seemingly consistent effective field theories need to obey, it also offers the opportunity to refine our understanding of
full string theory/quantum gravity
in various perturbative limits and most wanted  even in the interior of moduli space.

Indeed, a common theme in many swampland considerations is the focus on infinite distances in moduli spaces, motivated by the swampland distance conjecture \cite{Ooguri:2006in} which predicts that in every such limit one obtains infinite towers of exponentially light states signaling the breakdown of the  effective description. This was refined via the emergent
string conjecture in \cite{Lee:2019wij} saying that one can distinguish two typical such limits,
namely decompactification and emergent string limits.
Despite the ever growing number of swampland conjectures attempting to provide qualitative criteria for effective field theories by studying such limits, quantitative information remains limited. Studying the quantitative effects of the existence of infinite towers of light states is precisely the aim of the emergence proposal \cite{Palti:2019pca}, initially postulating that the kinetic terms for all light fields are emergent in the infrared by integrating out towers of light  states down from some ultraviolet scale below the Planck scale. This idea was initially explored in the context of field theoretic approximations \cite{Heidenreich:2017sim,Grimm:2018ohb,Castellano:2022bvr,Casas:2024ttx}, looking not only at kinetic terms but also effective potentials. More recently, it was observed that emergence can resonate nicely with certain black hole results in \cite{Calderon-Infante:2023uhz,Calderon-Infante:2025pls}. A subsequent extrapolation of the initial emergence proposal is to expect the entire low energy effective action to be emergent in the sense of arising via integrating out light states.

However, while it is natural, in view of string perturbation theory, to expect that only the light towers of states, which are perturbative from the point of view of emergence, should contribute to the emergence calculation, expecting them to always suffice to generate the entire effective action  in any infinite distance limit seems quite radical. 
Building on previous field theoretic results and taking under consideration the limitations of the emergence proposal in both emergent string \cite{Blumenhagen:2023yws} and decompactification limits \cite{Lee:2018urn}, an M-theoretic refinement of the emergence proposal was formulated in \cite{Blumenhagen:2024ydy,Blumenhagen:2024lmo}:
\begin{quotation}
\noindent
{\it  M-theoretic Emergence Proposal:
In the  infinite distance decompactification (M-theory) isotropic limit $M_* R_{11}\gg 1$ with the lower dimensional Planck scale kept fixed,  a perturbative quantum gravity theory arises whose low energy effective description emerges via quantum effects by integrating out the full infinite towers of states with a mass scale parametrically not larger than the eleven-dimensional Planck scale. 
These perturbative towers are transverse $M2$-, $M5$-branes carrying momentum along the eleventh direction ($D0$-branes) and along any potentially present compact direction.}
\end{quotation}
Firstly, this claims that in an isotropic decompactification limit to M-theory, approached while keeping the $d$-dimensional Planck scale fixed, a new quantum gravity theory arises whose low energy effective action is generated completely by quantum effects.
Secondly, these quantum effects are arising
by integrating out the full infinite towers of perturbative states, whose typical mass scale is not larger the eleven-dimensional Planck mass, which is the species scale in this
decompactification limit.
In the language of type IIA string theory these states are $D0$-branes and wrapped $D2$- and $N\!S5$- branes carrying Kaluza-Klein momentum. 
In addition there are ``non-perturbative'' states in this limit,
which are given by longitudinally wrapped  $M2$- and $M5$-branes.
These in principle will  also contribute to certain couplings,
but being thought of as coherent bound states  of the perturbative states,
should not be integrated out in addition. This is in the same spirit
as in perturbative string theory, where the $D$-branes are coherent
states, i.e. boundary states, of closed strings and are not integrated out
explicitly in string diagrams.

Noting that this proposal is in accord with
  the Matrix Model description of M-theory (see e.g. \cite{Banks:1996vh} and related reviews \cite{Bilal:1997fy,Bigatti:1997jy,Taylor:2001vb}),
it was concretely tested for simple amplitudes that
are protected by supersymmetry so that one only needed
the well known BPS spectrum of M-theory. It turned out
that in these examples already a one-loop Schwinger integral
involving the light towers of states sufficed to
obtain the full exact coupling. 
In \cite{Blumenhagen:2024ydy}, the $R^4$-coupling in maximally supersymmetric toroidal compactifications of M-theory and in \cite{Blumenhagen:2023tev} the holomorphic prepotential in a four dimensional (non-compact) Calabi-Yau compactification of type IIA could be determined via an emergence calculation. The latter approach motivated Hattab/Palti \cite{Hattab:2023moj,Hattab:2024thi,Hattab:2024chf,Hattab:2024yol,Hattab:2024ssg} to refine the methods
developed initially by Gopakumar/Vafa \cite{Gopakumar:1998ii,Gopakumar:1998jq} and to even
give a new complex contour integral representation of the prepotential (an alternative one was studied in \cite{Castellano:2025ljk}).

It is clear that the first step to perform an emergence calculation is identifying which are the states that need to be integrated out to give rise to a term in the effective action of the theory. In \cite{Blumenhagen:2024ydy,Blumenhagen:2024lmo} it was suggested that from the point of view of the emergence proposal the effective cutoff of a theory could also be interpreted as the energy scale distinguishing between non-perturbative and perturbative degrees of freedom.
 In the presence of light states the ultraviolet cutoff is no longer the $d$-dimensional Planck scale $M_{\rm pl}^{(d)}$, but rather a lower energy scale called the species scale \cite{Dvali:2007hz,Dvali:2007wp,Dvali:2010vm} (see also \cite{Veneziano:2001ah} for some earlier results), which is given by
\begin{equation}\label{species scale gen}
    \Lambda_{\rm sp}=\frac{M_{\rm{pl}}^{(d)}}{N_{\rm sp}^{1/(d-2)}}\,,
\end{equation}
where $N_{\rm sp}$ is the number of light states present in the theory. While this definition can safely be used in the case of particle towers, it naively results in multiplicative logarithmic corrections to the species scale when string towers are considered \cite{Marchesano:2022axe,Castellano:2022bvr,Blumenhagen:2023yws}, due to the exponential degeneracy of their spectrum. 

Apart from a pure field theoretic approach, it has been suggested that one should define the species scale as the energy scale suppressing higher curvature corrections to the Einstein-Hilbert action (up to finitely many fine tunings) \cite{vandeHeisteeg:2023dlw}, initially motivated by associating it to the one-loop topological free energy in four dimensions with $N=2$ supersymmetry \cite{vandeHeisteeg:2022btw}, but further tested in \cite{vandeHeisteeg:2023dlw,Castellano:2023aum}. This definition  may even be extended to the interior of the moduli space, including the so-called desert point \cite{Long:2021jlv}.\footnote{Gaining a deeper understanding of the various energy scales characterizing an effective theory of quantum gravity is an area of active research within the swampland community, an extensive review of which falls beyond the scope of this paper. For example, recently the species scale has been distinguished from other black hole scales \cite{Bedroya:2024uva} and a more detailed analysis of the relevant energy scales for the suppression of different operators in gravitational expansions has been carried out in \cite{Calderon-Infante:2025ldq}. Additionally, it has also been studied from a thermodynamics point of view in \cite{Cribiori:2023ffn,Basile:2024dqq,Herraez:2024kux}.  }  Motivated by these arguments, we will be ignoring the multiplicative corrections to the species scale in the case of light string towers being present, as modular invariance only allows for additive ones \cite{Cribiori:2023sch}.
On a practical level, multiple towers might be becoming light in infinite distances limits and the corresponding species scale can be calculated using the algorithm of \cite{Castellano:2021mmx}. To briefly summarise it, after identifying the lightest tower of states and calculating the species scale corresponding to it, we compare this scale with the next lightest tower and if this tower lies below it we have to include it in our calculations of the species scale. If one of these light towers is a string tower, it dominates and thus saturates the species scale.

This work aims to confront the M-theoretic emergence proposal, this time focusing on a gauge rather than a gravitational coupling, namely an $F^4$-coupling in six dimensions. We will be adopting a complementary approach to previous works by taking advantage of the mathematical methods already developed in order to explicitly show that the $D4$-branes present on the type IIA side
are indeed redundant, as expected from emergence.
This time, despite the fact that we are dealing with modular forms (Eisenstein series) similar to earlier works, we will be collecting evidence suggesting their degeneration behavior which to the best of our knowledge has not been tested on the level of the representations that we will be considering. As we will see, despite the technical subtleties of these calculations the emergence proposal can be a guiding principle in organizing the various terms in a way convenient to shed light on various non-trivial cancellations. Let us stress that our goal here is not only  to extend emergence considerations to yet another coupling, but to also support
the prerequisite of the M-theoretic emergence proposal, namely
the role of the heavy $D4$-branes in a geometrically simple setup, where they are computationally accessible with our methods.

In section \ref{sec2}, we will be reviewing the details of the heterotic/type IIA duality in six dimensions which we will be heavily relying on. The reason for this is twofold. On a conceptual level, this is an interesting instance where our M-theoretic results can be useful in a decompactification limit with weakly coupled strings present. On a practical level, we will see that it is much easier to keep track of the various physical contributions on the heterotic side while remaining very close to the analysis of \cite{Blumenhagen:2024ydy}. It is solely for this reason that the majority of calculations will be performed in terms of heterotic quantities, despite the M-theoretic emergence arguments that have motivated us. However, to turn the logic around,  it is also a peculiar instance where M-theoretic arguments can guide us into better understanding the microscopic structure of a string theoretic amplitude. We will also introduce the gauge coupling of interest in more detail. In section \ref{sec3}, we will proceed with a case-by-case study of all terms which could be receiving $D4$-brane contributions on the type IIA/M-theoretic side and deliver an almost complete proof of the redundancy of $D4$-brane degrees of freedom in this amplitude's description. In section \ref{sec4}, we will speculate on the extension of our calculation to five dimensions and wrapped $N\!S5$-branes before summarizing our findings in section \ref{sec5}.

\section{\texorpdfstring{$F^4$}{F\^{}4}-terms in six dimensions}
\label{sec2}
Recall that  most emergence calculations have been focusing on 1/2-BPS saturated quantities, which, due to supersymmetric protection, were one-loop exact.
We would now like to turn our attention to another such 1/2-BPS protected coupling, namely an $F^4$-coupling, in setups preserving 16 supercharges. We will focus on a very special six dimensional setup where the result has a group theoretic structure closely related to that of the $R^4$-term in six dimensions and where we can make use of triality relations associating type IIA and heterotic string theories.

\subsection{Duality and triality relations}

\begin{figure}[ht]
\centering

\includegraphics[width=0.8\textwidth]{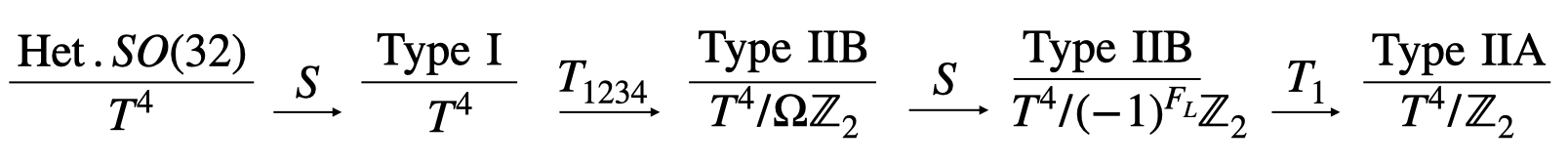}
\caption{Chain of maps between heterotic string theory compactified on $T^4$ and type IIA on K3. $T_i$ corresponds to a T-duality along the $i$-th direction and $S$ to an S-duality.}
\label{fig_duality map}
\end{figure}

Let us start by reviewing the string-string duality relations between heterotic and type IIA string theory in six dimensions. Heterotic string theory compactified on $T^4$ is conjectured to be dual to type IIA on K3 \cite{Sen:1995cj}. At the $T^4/\mathbb{Z}_2$ orbifold point of K3 this duality can be obtained by a chain of dualities (see e.g. \cite{Dabholkar:1997zd} for a detailed analysis)
shown in Figure \ref{fig_duality map},
that map the weak coupling regime of the heterotic $SO(32)$ string theory to the strongly-coupled type IIA theory in six dimensions. 

\noindent
Although the duality holds for any point in the moduli space of K3, we will be focusing on this orbifold limit as the corresponding conformal field theory description is completely solvable and we have a geometric interpretation of the particle states in terms of wrapped branes on the even cycles of $T^4$. 
Applying the chain of dualities we obtain
\begin{equation}
    \label{hetIIAmap}
    l_{\rm H}=g^{(6)}_{\rm IIA}\, l_{\rm IIA}\,,\qquad g^{(6)}_{\rm H}=\left(g^{(6)}_{ \rm IIA}\right)^{-1},\qquad \left(R_1^{\rm H}\right)^2:=\left(\frac{R_1}{l_{\rm H}}\right)^2=\frac{V_{\rm K3}}{l_{\rm IIA}^4}\,,
  \end{equation}
  where $g^{(6)}$ denotes the six-dimensional string coupling constants given by
  \begin{equation}
      \left(g^{(6)}_{\rm H}\right)^2=\frac{l_{\rm H}^ 4}{V_{T^4}}\, g_{\rm H}^2\,,\quad\qquad \left(g^{(6)}_{\rm IIA}\right)^2=\frac{l_{\rm IIA}^ 4}{V_{\rm K3}}\, g_{\rm IIA}^2\,.
  \end{equation}
  Note that the last T-duality from type IIB to type IIA distinguishes
  the first direction so that, in the end, the size of this direction
  on the heterotic side is mapped to the four-dimensional volume
on the type IIA side.

On the type IIA side, the $2$-cycles of $T^4/\mathbb{Z}_2$ get divided into six bulk cycles inherited from the $T^4$ and sixteen vanishing $2$-cycles at the $A_1$ singularities.
We will be considering a rectangular torus with   vanishing
Kalb-Ramond field restricted to the six 2-cycles of $T^4$.
Via the duality, this guarantees that also on the heterotic side
we have a rectangular torus with vanishing Kalb-Ramond field.
In this case, the duality map acts on the radii as
\eq{
\label{hetIIradiimap}
R_1^{\rm H}&=\sqrt{R_1^{\rm IIA}R_2^{\rm IIA}R_3^{\rm IIA}R_4^{\rm IIA}}\,, \qquad\qquad R_2^{\rm H}=\sqrt{\frac{R_3^{\rm IIA}R_4^{\rm IIA}}{R_1^{\rm IIA} R_2^{\rm IIA}}}\,,\\
R_3^{\rm H}&=\sqrt{\frac{R_2^{\rm IIA}R_4^{\rm IIA}}{R_1^{\rm IIA} R_3^{\rm IIA}}}\,,\qquad\qquad\phantom{aaaaaaa} R_4^{\rm H}=\sqrt{\frac{R_2^{\rm IIA}R_3^{\rm IIA}}{R_1^{\rm IIA} R_4^{\rm IIA}}}\,,
}
where these are given in string units.
Furthermore, the CFT $\mathbb Z_2$ orbifold  on the type IIA side
corresponds to a point in moduli space, where
there is half a unit of $B$-flux supported on all
the 16 fixed points, i.e. $\int_{S_i^2} B=1/2$ after
blowing up a fixed point to a finite size 2-sphere $S^2$.
Therefore, there will be an abelian gauge group $U(1)^{16}$
coming from the dimensional reduction
of the ten-dimensional R-R three-form $C_3$ on the sixteen  2-spheres
\eq{
         A_i =\int_{S_i^2} C_3\,.
}
On the heterotic side, this corresponds to a special choice
of discrete Wilson-lines along the $T^4$, also breaking
the initial non-abelian gauge group $SO(32)$ to its Cartan subalgebra.
The precise choice is not important for our purposes but can be found
in \cite{Kiritsis:2000zi}.

We will  focus on the 1/2-BPS saturated $F^4$-term for the diagonal linear combination 
of these sixteen gauge fields.
It was shown in \cite{Kiritsis:2000zi} that only these twisted sector gauge fields
have a non-vanishing $F^4$-coupling and that there is a precise
matching between the type IIA and the heterotic result.
For this to work, there are highly non-trivial cancellations happening such that the coupling
can be fully described by 1/2-BPS Kaluza-Klein and winding states on the heterotic side
\eq{
  A_{F^4}\sim l_{\rm H}^2\int {\frac{d^2\tau}{\tau_2^2}} \,  Z_{4,4}(R^{\rm H}_i) \sim
   l_{\rm H}^2 \,\mathcal{E}^{SO(4,4,\mathbb{Z})}_{\mathcal{V};1}\,,
 }
where    $Z_{4,4}(R^{\rm H}_i)$    denotes the 1/2-BPS partition function
of 1/2-BPS Kaluza-Klein and winding states
of the rectangular $T^4$ torus with vanishing B-field. 
This means that the contribution
of 1/2-BPS states involving any non-trivial right-moving string
excitation completely canceled out.
As indicated, this coupling can also be expressed in terms of a certain
Eisenstein series\footnote{Following the conventions of \cite{Obers:1999um}, the Eisenstein series of order $s$ for a representation $\mathcal{R}$ of a group $\mathcal{G}$ is $$
\mathcal{E}^{\mathcal{G}}_{\mathcal{R};s}=\hat{\sum_{m_i\in\mathbb{Z}}}\left[\sum\limits_{i,j}m_i\mathcal{M}_{ij}m_j\right]^{-s}=\frac{\pi^s}{\Gamma(s)}\hat{\sum_{m_i\in\mathbb{Z}}}\int_0^\infty\frac{dt}{t^{s+1}}e^{-\frac{\pi}{t}\sum\limits_{i,j}m_i\mathcal{M}_{ij}m_j}\,,$$
where $\mathcal{M}_{ij}$ transforms in $\mathcal{R}$ and
$\hat{\sum}_{m_i\in\mathbb{Z}}$ indicates that the term with all
$m_i=0$ has been excluded. 
} in the vector representation of $SO(4,4,\mathbb Z)$ 
\eq{
\label{Eisensteinvector}  \mathcal{E}_{\mathcal{V};1}^{SO(4,4,\mathbb{Z})}=\pi\hat{\sum\limits_{n_i,m_i}}\int_0^\infty\frac{dt}{t^2}\,\delta({\rm BPS})\,e^{-\frac{\pi}{t}\left(\sum\limits_i \left(n_i R_i^{\rm H}\right)^2+\sum\limits_i \left(m_i/R_i^{\rm H}\right)^{2}\right)}\,,
}
where the integers $m_i$ and $n_i$ denote the Kaluza-Klein and winding
modes along the $i$-th direction of the rectangular torus, respectively.  Hatted summations imply that setting all integers to zero is excluded.
The exponent is the mass square ${\cal M}^2_{\rm H}$ of the these modes in units
of the heterotic string scale $l_{\rm H}^{-2}$.
The BPS condition  takes the form of a homogeneous linear Diophantine equation  mixing winding and Kaluza-Klein modes \cite{Obers:1998fb}
\begin{equation}
    \label{BPSvectorrep}
    n_1\,  m_1+ n_2\, m_2+ n_3\, m_3+ n_4 \, m_4=0\,.
\end{equation}
Note that the expression \eqref{Eisensteinvector} is divergent and needs
to be regularized. Throughout this paper, for this purpose we invoke
the regularization method presented in \cite{Blumenhagen:2023tev,Blumenhagen:2024ydy}.

It was shown in \cite{Kiritsis:2000zi} that the tree-level coupling of  four twist fields on the type IIA side can be expressed
in a form that is rather typical for a genus one amplitude
\eq{
  \label{f4couplIIA}
  A_{F^4}\sim \frac{1}{g_{\rm IIA}^2}\frac{l_{\rm IIA}^6 g_{\rm IIA}^4}{V_{\rm K3}}\int {\frac{d^2\tau}{\tau_2^2}}\,   Z_{4,4}(R^{\rm IIA}_i) \,,
}
where $Z_{4,4}(R_i^{\rm IIA})$ is the 1/2-BPS the partition function including only the $D0$-, $D2$- and $D4$-branes  wrapped on even bulk cycles. Importantly, no 1/2-BPS states coming from the wrapping of $D2$-branes on the vanishing singularities is contributing.
The right hand side of \eqref{f4couplIIA} can also be 
expressed in terms of an Eisenstein series. Having in mind the M-theoretic emergence proposal, let us express it in terms of M-theory quantities. 
Indeed, we can describe strongly coupled type IIA on $T^4/\mathbb{Z}_2$ as M-theory on $S^1_{\rm M}\times T^4/\mathbb{Z}_2$, where $R_{11}$ is the radius of the M-theory circle. The type IIA quantities are related to the  M-theoretic ones through
\begin{equation}
    l_{\rm IIA}^2=(M_{*}^3R_{11})^{-1}\,,\qquad 
    g_{\rm IIA}^2=(M_*R_{11})^3\,,
\end{equation}
where $M_*=1/l_*$ denotes the eleven-dimensional Planck scale and $g_{\rm IIA}$ the ten-dimensional string coupling.
Then we can write
\eq{
      A_{F^4}\sim l_{\rm *}^2\,\mathcal{E}^{SO(4,4,\mathbb{Z})}_{\mathcal{C};1}=l_{\rm *}^2\, \mathcal{E}^{SO(4,4,\mathbb{Z})}_{\mathcal{S};1}\,,
}  
with the Eisenstein series in the conjugate spinor representation
\begin{equation}
\label{Eisensteincospinor}
\mathcal{E}_{\mathcal{C};1}^{SO(4,4,\mathbb{Z})}=\pi\hat{\sum\limits_{n_{ij},m,n_{D4}}}\int_0^\infty\frac{dt}{t^2}\,\delta({\rm BPS})\,e^{-\frac{\pi}{t}\left(\sum\limits_{i,j} (n_{ij}\, r_ir_j)^2+(m/r_{11})^{2}+(n_{D4}\, r_{11}\, v_{T^4})^2\right)}\,,
\end{equation}
where $r_i$ denote the radii  in units of $l_*$, and we abbreviated  $v_{T^4}=r_1r_2r_3r_4$.
Moreover, $n_{ij}$ denote the wrapping numbers of $M2$-branes along
the $(ij)$ 2-cycle of $T^4$,  $n_{D4}$ the wrapping number of longitudinal
  $M5$-branes, i.e. type IIA $D4$-branes, on the full $T^4$ and
  $m$ the Kaluza-Klein momentum along the 11-th direction.
  From the type IIA perspective, the latter is the number of $D0$-branes.
The exponent in \eqref{Eisensteincospinor} is the mass square ${\cal M}^2_{\rm M}$ of these modes in units of $l_*^{-2}$.
The 1/2-BPS condition of these wrapped branes reads
\begin{equation}
    \label{BPScospinor}
    n_{D4}\, m+n_{34}\, n_{12}+n_{24}\, n_{13}+n_{23}\, n_{14}=0\,.
\end{equation}

To sum up, heterotic/type IIA duality asserts that one has
the following relations for  the $F^4$-coupling
\begin{equation}\label{F^4allexpressions}
     A_{F^4}\sim l_{\rm H}^2\,\mathcal{E}^{SO(4,4,\mathbb{Z})}_{\mathcal{V};1}=l_{\rm H}^2\,\sum \frac{1}{\mathcal{M}_{\rm H}^2}=l_*^2\,\sum \frac{1}{\mathcal{M}_{\rm M}^2} =l_{\rm *}^2\,\mathcal{E}^{SO(4,4,\mathbb{Z})}_{\mathcal{C};1}=l_{\rm *}^2\, \mathcal{E}^{SO(4,4,\mathbb{Z})}_{\mathcal{S};1}\,.
\end{equation}
Note that the second equality is a direct consequence of the one-to-one mapping
of the 1/2-BPS states and their masses between the heterotic string and type IIA/M-theory.
However, on the level of a priori diverging  integral representations the validity of these expressions becomes apparent only after regularization and a subsequent
instanton expansion. One of the main objectives of this paper is precisely to
impose a recipe for how this expansion can be performed in practice.
The final equality  is a manifestation of the $SO(4,4)$ triality and was conjectured in \cite{Obers:1999um}.
We recall that the emergence proposal claims that the full tree-level type IIA
coupling is generated as a one-loop effect in M-theory, which is
already evident from its  representation in terms of the co-spinor
Eisenstein series, which can be viewed as a Schwinger
integral for integrating out the 1/2-BPS $D0$-$D2$-$D4$ bound states.

As already anticipated, the duality map \eqref{hetIIAmap} and \eqref{hetIIradiimap} matches the BPS spectra of both theories, i.e. 
it maps the vector to the conjugate spinor representation of the T-duality group $SO(4,4,\mathbb{Z})$. Specifically, it relates the charges of these representations as follows
\begin{equation}
(m_1;m_2,m_3,m_4;n_2,n_3,n_4;n_1)=(m;n_{12},n_{13},n_{14};n_{34},n_{42},n_{23};n_{D4})\,.
\end{equation}
Importantly, the $D0$-branes get mapped to Kaluza-Klein momenta along the first direction while the $D4$-branes wrapping the entire K3 correspond to winding modes along the same direction. The rest of the states, i.e. the $D2$-branes wrapping bulk $2$-cycles of the $T^4/\mathbb{Z}_2$, get mapped to Kaluza-Klein and winding modes along the other directions. Although our conventions for the map of $D2$-branes differ slightly from the ones in \cite{Kiritsis:2000zi} (c.f. \cite{Bergman:1999kq}), this will not affect our analysis.
Note that the duality maps the two 1/2-BPS conditions of the vector \eqref{BPSvectorrep}
and the co-spinor \eqref{BPScospinor} to each other.
The details of solving these types of equations are summarized in appendix \ref{app:Diophantine eqs} and are following the lines of \cite{Blumenhagen:2024ydy}.

\subsection{The decompactification limit}

In view of the emergence proposal for M-theory, the question now
is how these $F^4$-terms behave in the isotropic M-theory limit.
In this decompactification limit one takes $R_{11}\to \infty$ while
keeping the six-dimensional Planck scale and the  size of the K3
in M-theory units  constant.
Following \cite{Blumenhagen:2024ydy}, this limit is given by taking $\lambda\gg1$ and scaling
\begin{equation}
    R_{11}\rightarrow\lambda R_{11}\,,\quad M_*\rightarrow\lambda^{-\frac{1}{5}}M_*\,,\quad R_i\rightarrow \lambda^{\frac{1}{5}}R_i\,,
\end{equation}
where $R_i$ are the physical compactification radii.
Then, as anticipated, the six dimensional Planck scale $M_{\rm pl}^{(6)}$ does not scale
\begin{equation}
    \left( M_{\rm Pl}^{(6)}\right)^4=M_*^9\,V_{\rm K3}R_{11}\rightarrow \lambda^{-\frac{9}{5}}\lambda^{\frac{4}{5}}\lambda\, M_*^9\,V_{\rm K3}R_{11}=\left( M_{\rm Pl}^{(6)}\right)^4\,.
  \end{equation}
  
In terms of type IIA quantities this limit reads
\begin{equation}
  g_{\rm IIA}\rightarrow\lambda^{\frac{6}{5}}g_{\rm IIA}\,,\quad l_{\rm IIA}\rightarrow\lambda^{-\frac{1}{5}}l_{\rm IIA}\,,\quad g_{\rm IIA}^{(6)}\rightarrow\lambda^{\frac{2}{5}}g_{\rm IIA}^{(6)}\,, \quad R_i^{\rm IIA}\rightarrow  \lambda^{\frac{2}{5}}R_i^{\rm IIA}\,,
\end{equation}
so that it is clearly a co-scaled strong coupling limit.
One can check that this is an isotropic decompactification limit with the
Kaluza-Klein modes along the eleventh direction ($D0$-branes) being
the lightest tower of states.  The corresponding species scale
is then the seven-dimensional Planck scale
\eq{
  \label{speciessca}
  \Lambda_{\rm sp}\simeq M_{*} \, v_{T^4}^{1/5} \sim  M_{\rm Pl}^{(7)}\,,
}
which scales in the same way as the eleven-dimensional Planck scale.
Transversally wrapped $M2$-branes along the $(ij)$-plane have a mass $M\sim M_* r_i r_j$ so that
they scale in same manner as the species scale and hence,
following the M-theoretic refinement of the emergence proposal suggested in \cite{Blumenhagen:2024ydy}, 
should be considered as perturbative states in this limit.
It is important to note that longitudinally wrapped $M5$-branes ($D4$-branes) will be parametrically heavier than the species scale and therefore  should be considered as
non-perturbative objects in this limit. Hence they should
not be integrated out in the perturbative one-loop amplitude.

Using \eqref{hetIIAmap} and \eqref{hetIIradiimap} we can describe the corresponding limit in the dual heterotic frame
\begin{equation}
    g_{\rm H}\rightarrow g_{\rm H}\,,\quad l_{\rm H}\rightarrow\lambda^{\frac{1}{5}}l_{\rm H}\,,\quad g_{\rm H}^{(6)}\rightarrow\lambda^{-\frac{2}{5}}g_{\rm H}^{(6)}\,,\quad R_{2,3,4}^{\rm H}\rightarrow R_{2,3,4}^{\rm H}\,, \quad R_1^{\rm H}\rightarrow \lambda^{\frac{4}{5}} R_1^{\rm H}\,,
\end{equation}
where, interestingly, the ten-dimensional heterotic string coupling
as well as three out of the four  radii (in string units) of the $T^4$ do not scale.
Hence choosing $ g_{\rm H}\ll 1$,  the heterotic theory can stay in
the perturbative regime.  However, the radius of direction 1
grows large making the Kaluza-Klein modes along this direction the lightest
states in the theory. The species scale in this decompactification limit is
\eq{
  \Lambda_{\rm sp}\simeq \frac{1}{l_{\rm H}}  \left(\frac{R_2^{\rm H} R_3^{\rm H} R_4^{\rm H}}{g_{\rm H}^2}\right)^\frac{1}{5}\sim  M_{\rm Pl}^{(7)}\,,
}
which agrees of course with \eqref{speciessca} upon invoking the duality relations.
Note that the Kaluza-Klein and winding modes along directions 2,3,4 scale
in the same manner as the species scale and therefore should be considered
perturbative in the $\lambda\to\infty$ limit. However, the
winding modes along the first direction are parametrically heavier
and according to the emergence philosophy, should be
considered as non-perturbative states and  not be
integrated out in the heterotic one-loop amplitude.

Other decompactification limits can be considered, such as decompactifying a direction other than direction 1 on the heterotic side. In such cases, the corresponding type IIA dual does not yield an isotropic M-theory limit, but instead leads to an anisotropic limit in which two directions become large while the other two remain small. The lightest tower of states changes accordingly, i.e., in this limit, the lightest modes are $D2$-branes wrapping the two small directions. Computing the species scale shows that all states should be integrated out except the $D2$-branes wrapping the two large directions. On the heterotic side, this corresponds to integrating out all the states except the winding modes along the large circle. An analysis analogous to the one carried out in the next chapter can be done, obtaining the full result when all light states are integrated out. Nevertheless, we focus on the limit presented above, as any of these other limits can be obtained via a pair of T-dualities along direction 1 and another direction, so they do not really correspond to a physically different infinite distance limit.

It is the main objective of the following section to verify
that the contribution of winding modes along direction 1 to the $F^4$-coupling is indeed redundant.
In this respect, let us note that while for the Eisenstein series corresponding to the $R^4$-coupling studied in \cite{Blumenhagen:2024ydy} the emergence proposal could be viewed as a physical principle compatible with its mathematical properties, such properties have not been put forward for the Eisenstein series describing the $F^4$-coupling. It is, however, important to note that our calculations are focusing on vanishing axions, which would need to be implemented
if a more general relation was to be examined.

\section{Heterotic worldsheet instanton analysis}
\label{sec3}
As eluded to at the end of the previous section, the M-theoretic emergence proposal
suggests that in the respective decompactification limit
one does not have to integrate over the single tower of states
with a mass scale larger than the species scale.
For the $F^4$-term in  the heterotic theory this physical argument implies (at least for vanishing axions) the
mathematical relation\footnote{This relation is compatible with the decompactification limit of both Eisenstein series, see for example equation (C.14) in \cite{Green:2010wi}.}
\eq{
  \mathcal{E}^{SO(4,4,\mathbb{Z})}_{\mathcal{V};1}\sim
  \mathcal{E}^{SO(3,3,\mathbb{Z})}_{\mathcal{V}\oplus 1;1}\,,
}
where the representation of the Eisenstein series on the right hand side
is the direct sum of the  $SO(3,3,\mathbb{Z})$  vector representation
and a singlet. The latter is the singled out lightest Kaluza-Klein tower
along direction 1 with the corresponding winding mode
not summed over.

Following the method of \cite{Blumenhagen:2024ydy},
we will test this relation 
by carefully computing  the  worldsheet instanton contributions
on both sides.  Our strategy will be to start from the Eisenstein series in the vector representation on the heterotic side \eqref{Eisensteinvector}
or equivalently from  the co-spinor representation on the M-theory side \eqref{Eisensteincospinor} and apply Poisson resummations and the regularization scheme of \cite{Blumenhagen:2024ydy} to extract their instanton expansions.

In particular, this amounts to solving \eqref{BPSvectorrep} treating the winding modes as the coefficients and the Kaluza-Klein momenta as the variables. Upon mapping to the type IIA theory, this uniquely fixes a particular representation of the co-spinor Eisenstein series. Given the duality between winding modes along direction 1 and $D4$-brane windings, we will focus on the instantonic sector that arises by turning on at least one winding mode and investigate the role that winding modes along direction 1 play. The M-theoretic emergence proposal predicts that any terms related to $D4$-brane contributions should cancel each other.  We will neglect the pure Kaluza-Klein sector of the heterotic theory as they are not of interest for emergence because the winding modes along direction 1 are not present and no cancellations are possible at the level of the exponential expansion. Specifically, all the terms in the pure Kaluza-Klein sector are given in terms of the complex structure moduli of the $T^4$ or negative powers of the radii. Nevertheless, they can be computed with the methods of \cite{Blumenhagen:2024ydy}.

To make this presentation as self-contained as possible, we will review the approach of \cite{Blumenhagen:2024ydy} to solve the Schwinger-like integrals \eqref{Eisensteinvector} and \eqref{Eisensteincospinor} subject to the corresponding BPS conditions \eqref{BPSvectorrep} and \eqref{BPScospinor} respectively. 
Let us start by arguably the most complicated calculation on the heterotic side, which is that of \eqref{BPSvectorrep} with four non-vanishing winding numbers. Observe that this case was not treated in detail in \cite{Blumenhagen:2024ydy}, where perturbative terms were analyzed in dimensions $d\geq 7$.

Solving the Diophantine equation with the winding numbers as coefficients by following the steps of appendix \ref{app:Diophantine eqs}, we can re-express the Schwinger-like integral \eqref{Eisensteinvector} in terms of four coprime winding numbers denoted by $\mathbf{\tilde{n}_4}$ and three new unconstrained integers $\mu_i$ as
\begin{equation}
    A^{\rm H}_{4}=2^4\pi\,l_{\rm H}^2\sum_{N>0}\sum_{\mathbf{\tilde{n}_4}}{\sum\limits_{\mu_1,\mu_2,\mu_3\in\mathbb{Z}}}\int_0^\infty\frac{dt}{t^2}\,e^{-\frac{\pi}{t}\left(N^2L_{\rm H}^2+\mu_i\mathcal{M}^{ij}\mu_j\right)}\,,
\end{equation}
where $L_{\rm H}^2=\sum_{i=1}^4\left(\tilde{n}_iR_i^{\rm H}\right)^2$ and $\mathcal{M}^{ij}$ is a $3\times 3$ matrix given in \eqref{MMatrix5}. Applying a Poisson resummation\footnote{\label{footpoissonresum}
For a symmetric positive definite $k\times k$ matrix $G$ and a real
vector $b^I$, Poisson resummation amounts to $$\sum_{m^I\in
  \mathbb{Z}^k} e^{-\frac{\pi}{t}\sum\limits_{I,J}(m^I+b^I)G_{IJ}(m^J+b^J)} =
\frac{t^{\frac k2}}{\sqrt{\det(G_{IJ})}}\sum_{m_I \in\mathbb{Z}^k}
e^{-2\pi i\sum\limits_{I} m_I b^I - \pi t\sum\limits_{I,J} m_I G^{IJ}m_J}.$$} over the unconstrained integers $\mu_i$ we obtain
\begin{equation}\label{4windingsonintegral}
    A^{\rm H}_{4}=2^4\pi\,l_{\rm H}^2\,\sum_{N>0}\sum_{\mathbf{\tilde{n}_4}}{\sum\limits_{\mu_1,\mu_2,\mu_3\in\mathbb{Z}}}\int_0^\infty\frac{dt}{t^\frac{1}{2}}\frac{1}{\sqrt{{\rm det}(\mathcal{M})}}\,e^{-{\pi}{t}\,\mu_i\mathcal{M}_{ij}^{-1}\mu_j-\frac\pi tN^2L_{\rm H}^2}\,.
  \end{equation}
Note that  $\mu_1=\mu_2=\mu_3=0$  contributes to a constant term which will be studied independently in the following section so that we now assume that not all $\mu_i$ are vanishing.
Then, by recognizing that the appearing integrals  correspond to  integral representation of modified Bessel functions $K_\nu(x)$ of order $\nu$ \cite{Kiritsis:1997em}
\begin{equation}
\label{besselrel}
\int_0^\infty \frac{dx}{x^{1-\nu}} \,e^{-{\frac{b}{x}}-cx}=2 \left|  {\frac{b}{c}}\right|^{\frac{\nu}{2}} K_\nu\Big(2\sqrt{|b\, c|}\Big)\, ,
\end{equation}
it was observed already in \cite{Obers:1998fb} that they can be interpreted as instantonic contributions.
Combining the symmetry property of the Bessel functions $K_{-\nu}(x)=K_\nu(x)$ with the fact that $K_{1/2}(x)=\sqrt{\frac{\pi}{2x}} e^{-x}$ and the remarkably simple expression for the determinant of the matrix encoding the momentum contributions 
 det$(\mathcal{M})=L_{\rm H}^2/(V_4^{\rm H})^2$,  we get
\begin{equation}
    A^{\rm H}_{4}=2^4\pi\,l_{\rm H}^2V_4^{\rm H}\sum_{N>0}\sum_{\mathbf{\tilde{n}_4}}\hat{\sum\limits_{\mu_1,\mu_2,\mu_3\in\mathbb{Z}}}\frac{e^{-2\pi N L_{\rm H}\sqrt{\mu_i\mathcal{M}_{ij}^{-1}\mu_j}}}{L_{\rm H}\sqrt{\mu_i\mathcal{M}_{ij}^{-1}\mu_j}}\,,
\end{equation}
where $V_4^{\rm H}=R_1^{\rm H}R_2^{\rm H}R_3^{\rm H}R_4^{\rm H}$ is the $T^4$ volume in heterotic units. 

The key observation allowing us to extend the analysis of  \cite{Blumenhagen:2024ydy} is that we are able to express the exponent as the action of a bound state of at most six worldsheet instantons wrapping the different 2-cycles of $T^4$ in a form suitable for further calculations. Specifically, it is a non-trivial fact that the different coefficients can be arranged into perfect squares as follows 
\begin{equation}
    \label{vector rep final solution four wind modes}
    \left(L_{\rm H}\sqrt{\mu_i\mathcal{M}_{ij}^{-1}\mu_j}\right)^2=\sum_{i<j}\vartheta_{ij}^2 \,C_{ij}^2\,,
\end{equation}
with $\vartheta_{ij}=R_i^{\rm H}R_j^{\rm H}$ denoting  the area the instantons are wrapping, and $C_{ij}$ the respective coefficients. These are given in terms of particular solutions ($X_i,Y_i,Z_1$) of the Diophantine equations obtained iteratively from subsets of terms of \eqref{BPSvectorrep} (more details on the notation\footnote{When referring to a smaller set of coprime numbers obtained by the original ones by repeatedly dividing by various common divisors a hat symbol might be used. For example, we will usually have a set of four coprime numbers $\tilde{n}_1,\tilde{n}_2,\tilde{n}_3,\tilde{n}_4$, where $\tilde{n}_{1,2}=g_2g_3\hat{n}_{1,2}$, $\tilde{n}_3=g_3\hat{n}_3$ such that $\rm{gcd(}g_3,\tilde{n}_4)=1$, $\rm{gcd(}g_2,\hat{n}_3)=1$ with $g_2={\rm gcd(}\tilde{n}_1/g_3,\tilde{n}_2/g_3)$\,, $g_3={\rm gcd(}\tilde{n}_1,\tilde{n}_2,\tilde{n}_3)$ and $\rm gcd$ standing for greatest common divisor. 
\label{hatnotation}
} and properties of these solutions can be found in appendix \ref{app:Diophantine eqs}), the unconstrained integers $\mu_i$ and the winding numbers as  
\eq{\label{coefshet6instantons}
  C_{12}&=g_2g_3\mu_1\,,\quad\quad    C_{34}=(Y_0X_1-X_0Y_1)\mu_1+(\hat{n}_1X_1+\hat{n}_{2}Y_1)\mu_2+\hat{n}_3\mu_3\,,  \\ 
  C_{23}&=g_3(X_0\mu_1-\hat{n}_2\mu_2)\,,\quad\quad  C_{14}=(Y_0Z_1-g_2Y_1)\mu_1+\hat{n}_1Z_1\mu_2-\hat{n}_1g_2\mu_3\,,\\
   C_{24}&=-(X_0Z_1-g_2X_1)\mu_1+\hat{n}_2Z_1\mu_2-g_2\hat{n}_{2}\mu_3\,, \quad\quad  C_{13}=g_3(Y_0\mu_1+\hat{n}_1\mu_2)\,,
}
where $g_i$ denotes the greatest common divisor of $i$-winding numbers. 

Despite the complexity of these coefficients, one can prove that it is not possible to set more than three of them to zero consistently.
Another key aspect of \eqref{vector rep final solution four wind modes} is that the various coefficients satisfy the 1/2-BPS condition for the worldsheet instantons 
\begin{equation}\label{BPS analogue}
    C_{12}\,C_{34}+C_{14}\,C_{23}+C_{13}\,C_{24}=0\,.
\end{equation}
Having fixed the representation on the heterotic side, the type IIA representation can be read off directly from \eqref{vector rep final solution four wind modes} after using the relations \eqref{hetIIAmap} and \eqref{hetIIradiimap}. Concretely, this is the instanton contribution one gets from solving \eqref{BPScospinor} treating  $n_{D4},n_{34},n_{42},n_{23}$ as coefficients.
In that case, the matrix $\mathcal{N}$ encoding the contributions of the remaining $D2$-branes and $D0$-branes has a determinant ${\rm det}(\mathcal{N})=L_{\rm IIA}^2r_1^4r_{11}^{-2}$, resulting in
\begin{equation}
\label{cospinor rep final solution D4, 1 singled out}
    A^{\rm IIA}_{D4,n_{1j}\neq0}=2^4\pi\,\frac{l_*^2r_{11}}{r_1^2}\sum_{N>0}\sum_{\mathbf{\tilde{n}_4}}\hat{\sum\limits_{\mu_1,\mu_2,\mu_3\in\mathbb{Z}}}\frac{e^{-2\pi N L_{\rm IIA}\sqrt{\mu_i\mathcal{N}_{ij}^{-1}\mu_j}}}{L_{\rm IIA}\sqrt{\mu_i\mathcal{N}_{ij}^{-1}\mu_j}}\,,
\end{equation}
where 
\begin{equation}
    \left(L_{\rm IIA}\sqrt{\mu_i\mathcal{N}_{ij}^{-1}\mu_j}\right)^2=\sum_{\substack{1<i<j}} (t_{ij} r_{11})^2\, c_{ij}^2+\sum_{i=2}^4\Big(\frac{r_i}{r_1}\Big)^2c_{kl}^2\,,\quad k\neq l\neq i\neq 1\,,
  \end{equation}
where $t_{ij}=r_i r_j$ and $L_{\rm IIA}$ and $c_{ij}$ are given after applying the relations \ref{hetIIAmap} and \ref{hetIIradiimap} to the heterotic result and expressing the result in M-theory units. One can readily check our claim that the different expressions in \eqref{F^4allexpressions} match upon expanding the relevant integral representations. Moreover, note that we do not obtain every possible worldsheet instanton here, but with our chosen representation on the heterotic side the missing ones are dual to those given by the heterotic pure Kaluza-Klein sector, which, as mentioned before, are not relevant for emergence purposes. Let us emphasize that without the duality to the heterotic theory, choosing the most appropriate representation for a systematic study of the $D4$-brane contributions would be significantly more difficult.

Having made sure that the most complicated case of four non-trivial terms being present in our BPS conditions is manageable, let us also collect here the simpler solutions corresponding to less heterotic string winding modes being turned on. The methods used to obtain these contributions
are identical to the ones presented for the case of four non-vanishing winding numbers.

All four cases of three winding modes being non-zero give rise to similar contributions
\eq{\label{3windings het general}
    A_{n_l=0}^{\rm H}=&2\cdot 2^3\pi\,l_{\rm H}^2V_4^{\rm H}\sum_{\substack{N>0\\n_i,n_j,n_k>0\\m_i,m_j,m_k\\\rm{BPS}}}\frac{e^{-2\pi N\sqrt{L_{{\rm H},n_l=0}^2+m_i^2\vartheta_{jk}^2+m_j^2\vartheta_{ik}^2+m^2_k\vartheta_{ij}^2}}}{\sqrt{L_{{\rm H},n_l=0}^2+m_i^2\vartheta_{jk}^2+m_j^2\vartheta_{ik}^2+m^2_k\vartheta_{ij}^2}}\\
    &+2^3\pi\,l_{\rm H}^2 V_4^{\rm H}\sum_{N>0}\hat{\sum\limits_{m_i,m_j,m_k}}\sum_{\substack{\tilde{\mathbf{n}}_3\\ {\rm BPS} }}\frac{e^{-2\pi N\sqrt{m_i^2\vartheta_{jk}^2+m_j^2\vartheta_{ik}^2+m^2_k\vartheta_{ij}^2}}}{\sqrt{m_i^2\vartheta_{jk}^2+m_j^2\vartheta_{ik}^2+m^2_k\vartheta_{ij}^2}}\,,}  
where $L_{{\rm H},n_l=0}^2=\sum_{i\neq l}n_i^2\vartheta_{il}^2=\sum_{m_l>0}\sum_{i\neq l}({\tilde{n}_im_l})^2\vartheta_{il}^2$, with $\tilde{n}_3=(\tilde{n}_i,\tilde{n}_j,\tilde{n}_k)$ coprime. It is indicated that the integers are also subject to the BPS condition \eqref{BPSvectorrep}, this being the reason why this term cannot contribute to single instantons or bound states of four instantons. 

Two non-zero winding numbers can at most generate a bound state of five instantons
\begin{equation}\label{2windings het gen2}
    A^{\rm H}_{n_i,n_j\neq0}=2^2\pi\,l_{\rm H}^2V_4^{\rm H}\hspace{-1pt}\sum_{\substack{N>0\\\hat{n}_i,\hat{n}_j\\M,m_k,m_l\neq\vec{0}}}\hspace{-10pt}\frac{e^{-2\pi N\sqrt{M^2\vartheta_{ij}^2+\hat{n}_i^2m_k^2\vartheta_{ik}^2+\hat{n}_j^2m_k^2\vartheta_{jk}^2+\hat{n}_i^2m_l^2\vartheta_{il}^2+\hat{n}_j^2m_l^2\vartheta_{jl}^2}}}{\sqrt{M^2\vartheta_{ij}^2+\hat{n}_i^2m_k^2\vartheta_{ik}^2+\hat{n}_j^2m_k^2\vartheta_{jk}^2+\hat{n}_i^2m_l^2\vartheta_{il}^2+\hat{n}_j^2m_l^2\vartheta_{jl}^2}}\,,
\end{equation}
while a single non-zero winding number can give at most a bound state of three instantons
\begin{equation}
\label{1winding het gen}
  A^{\rm H}_{n_i\neq0}= 2\pi\,l_{\rm H}^2V_4^{\rm H}\sum_{n_i>0}\hat
  {\sum\limits_{m_j,m_k,m_l}}\frac{e^{-2\pi n_i\sqrt{m_j^2\vartheta_{ij}^2+m_k^2\vartheta_{ik}^2+m_l^2\vartheta_{il}^2}}}{\sqrt{m_j^2\vartheta_{ij}^2+m_k^2\vartheta_{ik}^2+m_l^2\vartheta_{il}^2}}\,.
\end{equation}
All of these expressions can be mapped to the type IIA side to track down their dual contributions to the $F^4$-amplitude via \eqref{hetIIradiimap}. 

All the results of the upcoming analysis can be obtained by careful manipulations of the previous formulas. To be more precise, we will analyze each possible bound state of worldsheet instantons by summing the contributions of different winding modes being turned on. Table \ref{table_contributions} lists the different possibilities, though note that depending on which directions the instantons wrap some of these contributions might  be absent.
\vspace{5pt}
\begin{table}[ht]
\renewcommand{\arraystretch}{1.5} 
    \begin{center}
    \begin{tabular}{|c|c|}
    \hline
    $\#$ winding numbers & $\#$ worldsheet instantons  \\
    \hline \hline 
    1& 1, 2, 3\\
    \hline 2& 1, 2, 3, 4, 5\\
    \hline 3& 2, 3, 5, 6\\
    \hline 4& 3, 4, 5, 6\\\hline
    \end{tabular}
    \caption{Bound states of worldsheet instantons for different number of winding numbers.}
    \label{table_contributions}
    \end{center}
\end{table}

Having summarized the main conceptual issues motivating our analysis as well as its computational background, we can now proceed with the more technical part of this work.  To summarize our findings, we will explicitly demonstrate the mutual cancellations of the contributions with $n_1\neq 0$ for the constant term and bound states of up to five worldsheet instantons and will check that such cancellations would also correspond to the expected result in the highly convoluted case corresponding to six instantons. This is a non-trivial confirmation of the M-theoretic emergence proposal, which showcases how it can lead to previously unknown relations even for well-studied string theoretic amplitudes.

\subsection{Regularized constant term}
\label{secconstantterm}

Let us recall that in most of the previous expressions we excluded the cases where all the Poisson resummed integers are zero.  After regularization these terms actually  contribute to a constant term, which we will now study separately. 
   
On the heterotic side, 
we are interested in collecting all of the constant terms that arise by imposing the BPS condition \eqref{BPSvectorrep} in \eqref{Eisensteinvector} and performing a Poisson-resummation
over $\mu_1,\mu_2,\mu_3$.
For that purpose, recall the expression \eqref{4windingsonintegral} and its generalization
for the case that some winding numbers do vanish. Then summing
over all possible winding numbers we arrive at
\begin{equation}
   A^{\rm H}= \pi\,l_{\rm H}^2\,\sum^4_{|\alpha|=1}\sum_{N>0}\sum_{\mathbf{\tilde{n}_\alpha}} 2^{|\alpha|}\!\! {\sum\limits_{\mu_1,\mu_2,\mu_3\in\mathbb{Z}}}\,\int_0^\infty\frac{dt}{t^\frac{1}{2}}\frac{1}{\sqrt{{\rm det}(\mathcal{M})}}\,e^{-{\pi}{t}\,\mu_i\mathcal{M}_{\alpha,ij}^{-1}\mu_j-\frac \pi tN^2L_{{\rm H},\alpha}^2}\,,
\end{equation}
where, following the notation of \cite{Blumenhagen:2024ydy}, $\mathbf{\alpha}$ is a vector with entries $\alpha_i\in\{0,1\}$ designating which of the winding numbers are non-vanishing. We denote $|\alpha|=\sum_i\alpha_i$ as the total number of non-vanishing winding numbers. After setting $\mu_1=\mu_2=\mu_3=0$, and using the fact that det$(\mathcal{M}_\alpha)=L_{{\rm H},\alpha}^2/(V_4^{\rm H})^2$ with $L_{{\rm H},\alpha}^2=\sum_{i=1}^4 \tilde{n}_i^2\left(R_i^{\rm H}\right)^2\alpha_i$, we obtain the constant term
\begin{equation}
    {\rm C}_{\rm H}=\pi\,l_{\rm H}^2V_4^{\rm H}\sum_{|\alpha|=1}^4\sum_{\mathbf{\tilde{n}}_\alpha}2^{|\alpha|}\frac{1}{L_{{\rm H},\alpha}}\sum_{N>0}\int_0^\infty\frac{dt}{t^{\frac 12}}e^{-\frac\pi tN^2L_{{\rm H},\alpha}^2}=\frac{\pi^2}{3}l_{\rm H}^2V_4^{\rm H}\,,
\end{equation}
where we made use of the fact that, since $L_{{\rm H},\alpha}$ cancels out, the various contributions only depend on the number of non-zero winding numbers $|\alpha|$ and the regularization of \cite{Blumenhagen:2023tev,Blumenhagen:2024ydy}. Let us recall  that this amounts to introducing an ultraviolet regulator for the integral, which is performed before  minimally subtracting divergencies due to this regulator and applying $\zeta$-function regularization of the final infinite sums. Here, one only needs the relation 
\begin{equation}\label{int exp/t^(1/2)}
\int_{\epsilon}^{\infty}\frac{d t}{t^{\frac 12}}\,e^{-\frac At}=\frac{2}{\sqrt\epsilon}-2\sqrt{\pi\,A}+\mathcal{O}(\sqrt{\epsilon})\,,
\end{equation}
as well as the fact that $\zeta(-1)=-1/12$, where $\zeta(s)=\sum_{n>0}n^{-s}$ is the Riemann $\zeta$-function. 

The question now is whether one does get the same result after leaving
out all contributions involving  heterotic winding modes along direction 1,
which are dual the type IIA $D4$-brane contributions. For illustrative purposes
we carry out this computation on the type IIA/M-theory  side.
Hence working in the co-spinor representation of the Eisenstein series, 
let us start by isolating the contributions arising from longitudinal   $M5$-branes, i.e. $D4$-branes.
Analogously to the heterotic matrix $\mathcal{M}_\alpha$,  we obtain the matrix $\mathcal{N}_\alpha$ with ${\rm det}(\mathcal{N}_\alpha)=(n_{D4}^2v_{T^4}^2r_{11}^2+L_{{\rm IIA},\alpha}^2)r_1^{4}r_{11}^{-2}$, where $\alpha$ is a 3-vector indicating which of the $D2$-brane wrapping numbers are non-vanishing and $L_{{\rm IIA},\alpha}^2=n_{34}^2(r_3r_4)^2\alpha_1+n_{42}^2(r_4r_2)^2\alpha_2+n_{23}^2(r_2r_3)^2\alpha_3$. This implies that after regularizing the integral as in \cite{Blumenhagen:2024ydy}, the total $D4$-brane contribution is given by
\begin{equation}
{\rm C}_{{\rm IIA},D4}=-2\pi^2\,\frac{l_*^2r_{11}}{r_1^2}\sum_{|\alpha|=0}^3 2^{|\alpha|+1}\sum_{\tilde{n}_{\alpha},n_{D4}\neq0}\hspace{-7pt}\zeta(-1)\,.
\end{equation}
Observe that here $|\alpha|$ could be zero because $n_{D4}\neq 0$. Again, thanks to the structure  of the determinant of $\mathcal{N}_\alpha$, the result only depends on the number of wrapped membranes present.
After using the regularization
\begin{equation}
\label{regcoprime2}
\sum_{\tilde{\mathbf n}_\alpha>0} 1\equiv \zeta(0)^{|\alpha|-1} = \left(-\frac12\right)^{|\alpha|-1}\,,
\end{equation}
we obtain\footnote{We have applied  the Binomial Theorem $(x+1)^k = \sum_{l=0}^k\binom{k}{l}x^l$ for $x=-1$.} 
 \begin{equation}
     {\rm C}_{{\rm IIA},D4}=-\frac{\pi^2}{3}\,\frac{l_*^2r_{11}}{r_1^2}\sum_{n_{D4}\neq 0}\sum_{|\alpha|=0}^3(-1)^{|\alpha|}\binom{3}{|\alpha|}=0\,.
 \end{equation}
Hence, as anticipated by the emergence proposal, the contributions from bound states involving $D4$ branes do indeed vanish.

Finally, let us check the constant term obtained from considering only the contributions from bound states of $D0$-branes and $D2$-branes and confirm that this matches the heterotic result. 
The BPS condition \eqref{BPScospinor} for $n_{D4}=0$ reduces to a linear Diophantine equation in three variables, leading to the introduction of two unconstrained integers, while the $D0$-brane charge $m$ is unrestricted. In a similar fashion to the previous case, after Poisson resumming over $\mu_1$, $\mu_2$ and $m$ and setting the Poisson resummed integers to zero, we get
\begin{equation}
    {\rm C}_{{\rm IIA},\cancel{D4}}=\pi\,\frac{l_*^2 r_{11}}{r_1^2}\sum_{|\alpha|=1}^3\sum_{\mathbf{\tilde{n}_\alpha}}\sum_{N>0}2^{|\alpha|}\frac{1}{L_{{\rm IIA},\alpha}}\int_0^\infty\frac{dt}{t^{1/2}}\,e^{-\frac{\pi}{t}N^2  L_{{\rm IIA},\alpha}^2}=\frac{\pi^2}{3}\frac{l_*^2 r_{11}}{r_1^2}\,.
\end{equation}
After mapping the heterotic result to type IIA units through \eqref{hetIIAmap} and \eqref{hetIIradiimap} and then to M-theory units, we confirm that the results match, as expected.

\subsection{Single heterotic instanton contributions}

From Table \ref{table_contributions} we infer that 
a single worldsheet  instanton contribution $(E\!F1)_{ij}$ can  arise from the following
winding number configurations
\begin{enumerate}
    \item[\hspace{1.5em}- Single winding mode:] $n_i\neq0$ or $n_j\neq0$\,,
    \item[\hspace{1.5em}- Two winding modes:]\, $n_i,n_j\neq0$\,.
\end{enumerate}
 In the first case, starting from \eqref{1winding het gen} and setting two of the momenta to zero, we obtain two identical contributions, so that we find in total 
\begin{equation}
   A_{(E\!F1_{ij}),1}^{\rm H}=2\cdot 2^2\pi\,l_{\rm H}^2V_4^{\rm H}\sum_{n>0}\sum_{m>0}\frac{e^{-2\pi n{m\vartheta_{ij}}}}{m\vartheta_{ij}}\,.
\end{equation}
For the second case, using \eqref{2windings het gen2}, setting $m_l=m_k=0$ and applying \eqref{regcoprime2}, we get
\eq{
A_{(E\!F1_{ij}),2}^{\rm H}&=2^3\pi\,l_{\rm H}^2V_4^{\rm H}\sum_{\substack{N,M>0\\\tilde{n}_i,\tilde{n}_j}}\frac{e^{-2\pi N\sqrt{M^2\vartheta_{ij}^2}}}{\sqrt{M^2\vartheta_{ij}^2}}=-2^2\pi\,l_{\rm H}^2V_4^{\rm H}\sum_{N,M>0}\frac{e^{-2\pi N{M\vartheta_{ij}}}}{{M\vartheta_{ij}}}\,.
}
To keep track of the physical origin of the various identical terms, we can schematically express the total contribution as
\begin{equation}
    A^{\rm H}_{E\!F1_{ij}}=A[n_i]+A[n_j]-A[n_i,n_j]\,,
\end{equation}
where we denote in brackets the instanton expressions given by the different winding modes turned on. Effectively, the final result is given by a single contribution of the first case. We distinguish two cases:
\begin{itemize}
    \item[-] If $i\neq 1\neq j$, then it has no obvious interpretation for emergence.
    \item[-] If $i=1$ (analogously $j=1$), then we can interpret this result as $n_1\neq 0$ contributions canceling out and the total contribution is given by the term $[n_j]$ ($[n_i]$). 
\end{itemize}
In both cases we can write the total result as
\begin{equation}
    A^{\rm H}_{E\!F1_{ij}}=2\pi\,l_{\rm H}^2V_4^{\rm H}\sum_{\substack{N>0 \\C_{ij}\neq 0}}\frac{e^{-2\pi N{|C_{ij}|\vartheta_{ij}}}}{|C_{ij}|\vartheta_{ij}}\,.
\end{equation}
Mapping this result to the type IIA side, we can confirm that obtaining the full result on the level of single instantons does not require including any $D4$-brane contributions.

\subsection{Double heterotic instanton contributions}

One can convince oneself that  the only possible BPS bound state of two worldsheet instantons is that of both wrapping a common direction, in other words $(E\!F1_{ij},E\!F1_{ik})$. This  could also be  predicted by \eqref{BPS analogue}.
From Table \ref{table_contributions}, for these states only the following contributions are relevant
\begin{enumerate}
    \item[\hspace{1.5em}- Three winding modes:] $n_i,n_j,n_k\neq0$\,,
    \item[\hspace{1.5em}- Two winding modes:] $n_j,n_k\neq 0$\,,
    \item[\hspace{1.5em}- Single winding mode:] $n_i\ne 0$\,.
\end{enumerate}
The first case corresponds to the second term of \eqref{3windings het general}, where we have set $l\neq i,j,k$. After taking $m_i=0$ and using \cite{Blumenhagen:2024ydy}
\begin{equation}
     \sum_{\substack{\tilde{\mathbf{n}}_3\\
      n_i m_i+n_jm_j=0\\ m_i,m_j\ne 0 }}  \!\!\!\!\!   1 =-\frac{1}{4}\,,
\end{equation}
we get
\begin{equation}
    A_{(E\!F1_{ij},E\!F1_{ik}),1}^{\rm H}=-2^3\pi\,l_{\rm H}^2V_4^{\rm H}\sum_{N>0}\sum_{m_j,m_k>0}\frac{e^{-2\pi N\sqrt{m_j^2\vartheta_{ik}^2+m_k^2\vartheta_{ij}^2}}}{\sqrt{m_j^2\vartheta_{ik}^2+m_k^2\vartheta_{ij}^2}}\,.
  \end{equation}
  The second case corresponds to \eqref{2windings het gen2} after setting $M$ and $m_l$ to zero and reabsorbing the remaining $m_i$ into $\hat{n}_j$ and $\hat{n}_k$ to get two unconstrained natural numbers
\begin{equation}
    A^{\rm H}_{(E\!F1_{ij},E\!F1_{ik}),2}= 2^3\pi\,l_{\rm H}^2V_4^{\rm H}\sum_{N>0}\sum_{n_j,n_k>0}\frac{e^{-2\pi N\sqrt{n_j^2\vartheta_{ij}^2+n_k^2\vartheta_{ik}^2}}}{\sqrt{n_j^2\vartheta_{ij}^2+n_k^2\vartheta_{ik}^2}}\,.
  \end{equation}
An identical contribution $ A^{\rm H}_{(E\!F1_{ij},E\!F1_{ik}),3}$ is found for the last case by setting $m_l=0$ in \eqref{1winding het gen}. Adding all these contributions, we get
\begin{equation}
    A_{(E\!F1_{ij},E\!F1_{ik})}^{\rm H}=A[n_i]+A[n_j,n_k]-A[n_i,n_j,n_k]\,.
\end{equation}
Again, we can distinguish two cases:
\begin{itemize}
    \item[-] If $i=1$, then the first and third contributions cancel each other out so there are no non-trivial contributions coming from $n_1\neq 0$.
    \item[-] If $i\neq 1$, then there are two options, either $j\neq 1\neq k$, so the different cancellations do not have an interpretation in terms of emergence, or $j=1$ (analogously $k=1$) and then the second and third contributions cancel each other out so that the final result is given by the term with $n_1=0$.
\end{itemize}
Overall, the double heterotic instanton contribution reads
\begin{equation}
    A_{(E\!F1_{ij},E\!F1_{ik})}^{\rm H}=2\pi\,l_{\rm H}^2V_4^{\rm H}\sum_{\substack{N>0 \\C_{ij},C_{ik}\neq 0}}\frac{e^{-2\pi N\sqrt{C_{ij}^2\vartheta_{ij}^2+C_{ik}^2\vartheta_{ik}^2}}}{\sqrt{C_{ij}^2\vartheta_{ij}^2+C_{ik}^2\vartheta_{ik}^2}}\,.
\end{equation}
By translating into type IIA/M-theory units, we can deduce that $D4$-branes are not contributing to double instanton contributions whatsoever, in agreement with the M-theoretic emergence proposal.

\subsection{Triple heterotic instanton contributions}

The various three instanton terms that can be constructed fall into two categories, namely either that the three instantons  are wrapping a common direction or that from the three instantons none is wrapping a particular direction. We can collectively denote these as $(E\!F1_{ij}, E\!F1_{ik}, E\!F1_{il})$ and $(E\!F1_{jk}, E\!F1_{jl}, E\!F1_{kl})$ respectively. Again this could have been inferred from \eqref{BPS analogue}. 

Let us first consider the case of all of them wrapping the $i$-direction. One can check that there are only three possible contributions to these states on the heterotic side, namely
\begin{enumerate}
    \item[\hspace{1.5em}-  Single winding mode:] $n_i\neq 0$\,,
    \item[\hspace{1.5em}- Three winding modes:] $n_i=0,n_j,n_k,n_l\neq0$\,,
    \item[\hspace{1.5em}- Four winding modes:] $n_1,n_2,n_3,n_4\neq 0$\,.
\end{enumerate}
For the first case, we consider \eqref{1winding het gen} and get
\begin{equation}
    A^{\rm H}_{E\!F1_{i(jkl)},1}=2^4\pi\,l_{\rm H}^2V_4^{\rm H}\sum_{\substack{n_i>0\\m_j,m_k,m_l>0}}\frac{e^{-2\pi n_i\sqrt{m_j^2\vartheta_{ij}^2+m_k^2\vartheta_{ik}^2+m_l^2\vartheta_{il}^2}}}{\sqrt{m_j^2\vartheta_{ij}^2+m_k^2\vartheta_{ik}^2+m_l^2\vartheta_{il}^2}}\,.
\end{equation}
For the second case, using the first term from \eqref{3windings het general} and setting all momenta to zero, we find
\begin{equation}
      A^{\rm H}_{E\!F1_{i(jkl)},2}=2^4\pi\,l_{\rm H}^2V_4^{\rm H}\sum_{n_j,n_k,n_l>0}\sum_{N>0}\frac{e^{-2\pi N\sqrt{n_j^2\vartheta_{ij}^2+n_k^2\vartheta_{ik}^2+n_l^2\vartheta_{il}^2}}}{\sqrt{n_j^2\vartheta_{ij}^2+n_k^2\vartheta_{ik}^2+n_l^2\vartheta_{il}^2}}\,.
\end{equation}
For the remaining case, we need to consider the contribution of \eqref{vector rep final solution four wind modes} when setting the prefactors of other instanton terms to zero. For simplicity, we will focus on the particular example $(E\!F1_{12},E\!F1_{13},E\!F1_{14})$ but the other cases would be treated the same way after reorganizing the BPS condition \eqref{BPSvectorrep}. Setting some of the prefactors to zero restricts the values of our previously unconstrained integers $\mu_i$. Despite the complexity of the general solution, the conditions of vanishing prefactors turn out to be a simple system of Diophantine equations satisfied by
\begin{equation}
    (\mu_1,\mu_2,\mu_3)=Q\left(\hat{n}_2,X_0,X_1\right)\,,\quad Q\neq0\,.
\end{equation}
Under these identifications, the remaining factors are
\begin{equation}
     \left(L_{\rm H}\sqrt{\mu_i\mathcal{M}_{ij}^{-1}\mu_j}\,\right)^2=Q^2\Big((g_2g_3\hat{n}_2)^2\vartheta_{23}^2+\hat{n}_3^2g_3^2\vartheta_{13}^2+\hat{n}_4^2\vartheta_{14}^2\Big)\,.
\end{equation}
Since ${\rm gcd}(\hat{n}_4,g_3\hat{n}_3,g_2g_3\hat{n}_2)={\rm gcd}(\hat{n}_4,g_3)=1$, we have three coprime numbers and multiplying them with $Q$ will give rise to an unrestricted set of four integers (just by exchanging the role of $N$ and $Q$), so that, after appropriate redefinitions of integers, we obtain
\begin{equation}
       A^{\rm H}_{(E\!F1_{1(234)}),3}=2^5\pi\,l_{\rm H}^2 V_4^{\rm H}\sum_{\substack{N>0\\n_1,n_2,n_3,n_4>0}}\frac{e^{-2\pi N\sqrt{{n}_2^2\vartheta_{12}^2+{n}_3^2\vartheta_{13}^2+n_4^2\vartheta_{14}^2}}}{\sqrt{n_2^2\vartheta_{12}^2+n_3^2\vartheta_{13}^2+n_4^2\vartheta_{14}^2}}\,,
\end{equation}
where an additional factor of $2$ arises by restricting the sum over $Q$ to natural numbers before absorbing it to go from a summation over coprimes to a summation over natural numbers. Since ${n}_1$ completely cancels out and $\zeta(0)=-1/2$, this contribution reduces to 
\begin{equation}
       A^{\rm H}_{(E\!F1_{1(234)}),3}=-2^4\pi\,l_{\rm H}^2 V_4^{\rm H}\sum_{\substack{N>0\\n_2,n_3,n_4> 0}}\frac{e^{-2\pi N\sqrt{{n}_2^2\vartheta_{12}^2+{n}_3^2\vartheta_{13}^2+n_4^2\vartheta_{14}^2}}}{\sqrt{n_2^2\vartheta_{12}^2+n_3^2\vartheta_{13}^2+n_4^2\vartheta_{14}^2}}\,.
\end{equation}
Thus for the general case,  the result is schematically given by
\begin{equation}
    A_{(E\!F1_{ij}, E\!F1_{ik}, E\!F1_{il})}^{\rm H}=A[n_i]+A[n_j,n_k,n_l]-A[n_1,n_2,n_3,n_4]\,.
\end{equation}
In order to interpret this in terms of emergence let us consider two cases:
\begin{itemize}
    \item[-] If $i=1$, then the first and third contributions cancel each other out so that effectively there is no contribution coming from $n_1\neq 0$.
    \item[-] If $i\neq 1$, then the second and third term are the ones canceling out and again the terms involving $n_1\neq 0$ do not contribute to the final result.
\end{itemize}
Hence, the total contribution for this case is given by
\begin{equation}
    A_{(E\!F1_{ij}, E\!F1_{ik}, E\!F1_{il})}^{\rm H}=2\pi\,l_{\rm H}^2V_4^{\rm H}\sum_{\substack{N>0 \\ C_{ij},C_{ik},C_{il}\neq0}}\frac{e^{-2\pi N\sqrt{C_{ij}^2\vartheta_{ij}^2+C_{ik}^2\vartheta_{ik}^2+C_{il}^2\vartheta_{il}^2}}}{\sqrt{C_{ij}^2\vartheta_{ij}^2+C_{ik}^2\vartheta_{ik}^2+C_{il}^2\vartheta_{il}^2}}\,.
\end{equation}

Let us now consider an example from the second category, namely  a bound state of three instantons none wrapping one direction. For example, for $(E\!F1_{13},E\!F1_{14},E\!F1_{34})$
contributions can arise from the following cases
\begin{enumerate}
    \item[\hspace{1.5em}- Three winding modes:] $n_2=0$\,,
    \item[\hspace{1.5em}- Two winding modes:] $n_1,n_3\neq 0$, or $n_1,n_4\neq0$, or $n_3,n_4\neq0$\,.
\end{enumerate}
The first case corresponds to the second term of \eqref{3windings het general} with  $l=2$ and forbidding any of the momenta to be zero, where we can eventually also use the regularization \cite{Blumenhagen:2024ydy}
\begin{equation}
\sum_{\substack{\tilde{\mathbf{n}}_3\\ \sum_{i=1}^3  n_i m_i=0\\ m_i\ne 0 }}  \!\!\!\!\!   1 =-\frac{1}{2}\,.
\end{equation}
For the second case(s) we will use \eqref{2windings het gen2}. After setting one of the $m's$ to zero and some trivial manipulations, the remaining prefactor is $2\cdot(1/2)^2=1/2$. In a similar way we would obtain the contributions for bound states of three instantons not involving directions 3 or 4. Then, for a bound state of three instantons not wrapping the $i$-th direction the result is
\begin{equation}
    A_{(E\!F1_{jk}, E\!F1_{jl}, E\!F1_{kl})}^{\rm H}=A[n_j,n_k]+A[n_j,n_l]+A[n_k,n_l]-2 A[n_j,n_k,n_l]\,.
\end{equation}
Again, we consider two cases:
\begin{itemize}
    \item[-] If $i=1$, then it has no significance for emergence.
    \item[-] If $i\neq1$, then $j=1$ (analogously $k=1$ or $l=1$) and the first and second term (respectively first and third terms or second and third terms) cancel the contribution of the fourth term, so that the total contribution involves no term with $n_1\neq 0$.
\end{itemize}
For the general case of the bound state of three instantons not wrapping the $i$-th direction, the total contribution reads
\begin{equation}
    A_{(E\!F1_{jk}, E\!F1_{jl}, E\!F1_{kl})}^{\rm H}=2\pi\,l_{\rm H}^2V_4^{\rm H}\sum_{\substack{N>0 \\C_{jk},C_{jl},C_{kl}\neq0}}\frac{e^{-2\pi N\sqrt{C_{jk}^2\vartheta_{jk}^2+C_{jl}^2\vartheta_{jl}^2+C_{kl}^2\vartheta_{kl}^2}}}{\sqrt{C_{jk}^2\vartheta_{jk}^2+C_{jl}^2\vartheta_{jl}^2+C_{kl}^2\vartheta_{kl}^2}}\,.
\end{equation}

To summarize, all the triple instanton contributions arising from the winding sector of heterotic string theory follow the pattern predicted by emergence, which is that the states corresponding to $n_1\neq0$ are canceling out.
Upon mapping the various contributions to the equivalent ones on the type IIA  side we confirm our expectation from emergence that $D4$-branes are redundant for the calculation of the amplitude.

\subsection{Four heterotic instanton contributions}

A priori one might naively guess that all instanton configurations are allowed,
but a careful inspection shows that, for example, the bound state  $(E\!F1_{12},E\!F1_{13},E\!F1_{14},E\!F1_{24})$ is completely absent. It is easy to see that such a term cannot be generated by turning on less than four winding modes and that when attempting to set the coefficients $C_{23}$ and $C_{34}$ to $0$ in \eqref{vector rep final solution four wind modes}, additional terms become trivial as well. In fact, one can similarly rule out any four instantons bound state with an analogous structure, i.e. three instantons wrapping a common direction and an additional term. This is also evident from \eqref{BPS analogue} for the case of four non-zero winding numbers.

The only allowed bound states will be thus of the form $(E\!F1_{ik},E\!F1_{il},E\!F1_{jk},E\!F1_{jl})$, where $i\neq j\neq k\neq l$. These only receive contributions from two cases
\begin{enumerate}
    \item[\hspace{1.5em}- Four winding modes:] $n_1,n_2,n_3,n_4\neq 0$\,,
    \item[\hspace{1.5em}- Two winding modes:] $n_i,n_j\neq 0$ or $n_k,n_l\neq 0$\,.
\end{enumerate}
For simplicity we will consider the particular example $(E\!F1_{13},E\!F1_{14},E\!F1_{23},E\!F1_{24})$, but the general case can be treated in the same way after reordering the coefficients of the BPS condition and repeating the same process.

Despite the intricate structure of \eqref{vector rep final solution four wind modes}, we are able to isolate this configuration by setting
\begin{equation}
(\mu_1,\mu_2,\mu_3)=Q(0,-\hat{n}_3,(\hat{n}_1X_1+\hat{n}_2Y_1))\,,\quad Q\neq0\,.
\end{equation}
We then get the following contribution for this particular bound state of instantons
\begin{equation}
\label{four instanton first case}
    A^{\rm H}_{(4EF),1}=2^5\pi\,l_{\rm H}^2 V_4^{\rm H}\sum_{\substack{\mathbf{\tilde{n}}_4\\Q,N>0}}\frac{e^{-2\pi N \sqrt{(\hat{n}_1^2(R_1^{\rm H})^2+\hat{n}_2^2(R_2^{\rm H})^2)(Q^2g_3^2\hat{n}_3^2(R_3^{\rm H})^2+Q^2\hat{n}_4^2(R_4^{\rm H})^2)}}}{\sqrt{(\hat{n}_1^2(R_1^{\rm H})^2+\hat{n}_2^2(R_2^{\rm H})^2)(Q^2g_3^2\hat{n}_3^2(R_3^{\rm H})^2+Q^2\hat{n}_4^2(R_4^{\rm H})^2)}}\,.
\end{equation}
It will prove convenient to manipulate the sums over this particular summand (denoted by $\ldots$ for convenience below) in the following way
\eq{\sum_{Q>0}\sum_{\mathbf{\tilde{n}_4}}\ldots&=\sum_{Q>0}\sum_{\substack{\hat{n}_4,g_3\\(\hat{n}_4,g_3)=1}}\sum_{\substack{\hat{n}_3,g_2\\(\hat{n}_3,g_2)=1}}\sum_{\tilde{n}_1,\tilde{n}_2}\ldots\\[0.1cm]
&=\hspace{-5pt}\sum_{n_4,Qg_3}\sum_{\substack{\hat{n}_3,g_2\\(\hat{n}_3,g_2)=1}}\sum_{\tilde{n}_1,\tilde{n}_2}\ldots=\hspace{-3pt}\sum_{\substack{Qg_3g_2 >0\\ n_4,n_3}}\sum_{\tilde{n}_1,\tilde{n}_2}\ldots\,.
}
In the first step we have written out the summation over the four coprime numbers in the hatted notation (see footnote\footref{hatnotation}), in the second step we have absorbed $Q$ so that $n_4=Q\hat{n}_4$ and we get a sum over two unconstrained natural numbers, and in the last step we have repeated this procedure for $n_3=\hat{n}_3 g_3Q$. It is important to mention that this can be done as the summand of \eqref{four instanton first case} depends only on these new variables. Furthermore, we get a sum over an unconstrained natural number given by the product $Qg_3g_2$ that does not appear in \eqref{four instanton first case}, which will result in
a $\zeta(0)=-1/2$ factor, leading to the result
\begin{equation}
    A^{\rm H}_{(4E\!F),1}=-2^4\pi\,l_{\rm H}^2 V_4^{\rm H}\sum_{\substack{N>0\\n_3,n_4>0}}\sum_{\hat{n}_1,\hat{n}_2}\frac{e^{-2\pi N \sqrt{\hat{n}_1^2{n}_3^2\vartheta_{13}^2+\hat{n}_2^2{n}_3^2\vartheta_{23}^2+\hat{n}_1^2{n}_4^2\vartheta_{14}^2+\hat{n}_2^2{n}_4^2\vartheta_{24}^2}}}{\sqrt{\hat{n}_1^2{n}_3^2\vartheta_{13}^2+\hat{n}_2^2{n}_3^2\vartheta_{23}^2+\hat{n}_1^2{n}_4^2\vartheta_{14}^2+\hat{n}_2^2{n}_4^2\vartheta_{24}^2}}\,.
\end{equation}

For the second case(s) we use \eqref{2windings het gen2} setting $M$ to zero and obtain
\eq{
    A^{\rm H}_{(4E\!F),2} =2\cdot 2^4\pi\,l_{\rm H}^2 V_4^{\rm H}\sum_{\substack{N>0,\\m_3,m_4>0}}\sum_{\hat{n}_1,\hat{n}_2}\frac{e^{-2\pi N \sqrt{\hat{n}_1^2{m}_3^2\vartheta_{13}^2+\hat{n}_2^2{m}_3^2\vartheta_{23}^2+\hat{n}_1^2{m}_4^2\vartheta_{14}^2+\hat{n}_2^2{m}_4^2\vartheta_{24}^2}}}{\sqrt{\hat{n}_1^2{m}_3^2\vartheta_{13}^2+\hat{n}_2^2{m}_3^2\vartheta_{23}^2+\hat{n}_1^2{m}_4^2\vartheta_{14}^2+\hat{n}_2^2{m}_4^2\vartheta_{24}^2}}\,,}
where one contribution is from the $n_i,n_j\neq 0$ sector and one from the  $n_k,n_l\neq 0$ sector.

The total term for the bound state of four instantons is schematically
\begin{equation}
    A_{(E\!F1_{ik},E\!F1_{il},E\!F1_{jk},E\!F1_{jl})}^{\rm H}=A[n_i,n_j]+A[n_k,n_l]-A[n_1,n_2,n_3,n_4]\,.
\end{equation}
Hence, indeed the two contributions involving $n_1$ cancel each other as predicted.  
We can express the total contribution of this bound state of instantons as
\begin{equation}
    A_{(E\!F1_{ik},E\!F1_{il},E\!F1_{jk},E\!F1_{jl})}^{\rm H}=2\pi\,l_{\rm H}^2 V_4^{\rm H}\hspace{-17pt}\sum_{\substack{N>0\\C_{ik},C_{il},C_{jk},C_{jl}\neq 0\\ {\rm BPS}}}\frac{e^{-2\pi N \sqrt{C_{ik}^2\vartheta_{ik}^2+C_{il}^2\vartheta_{il}^2+C_{jk}^2\vartheta_{jk}^2+C_{jl}^2\vartheta_{jl}^2}}}{\sqrt{C_{ik}^2\vartheta_{ik}^2+C_{il}^2\vartheta_{il}^2+C_{jk}^2\vartheta_{jk}^2+C_{jl}^2\vartheta_{jl}^2}}\,,
\end{equation}
where the coefficients of the worldsheet instantons are subject to the BPS condition \ref{BPS analogue}.
Analogously to the previous cases, upon mapping this result to the type IIA side we confirm that states involving $D4$-branes are not contributing to the final four instanton contribution to the amplitude.

\subsection{Five heterotic instanton contributions}

It is already evident from the four instanton analysis that increasing the number of instantons comprising our bound states involves more intricate relations between the appearing summations. This is also the case for the five instanton bound states, which we can, however, treat in generality. Let us comment again that in treating specific examples it might be convenient to rearrange the terms in the BPS conditions \eqref{BPSvectorrep} and \eqref{BPScospinor}.

Let us consider the five instanton bound state on the heterotic side corresponding to only instantons wrapping $\vartheta_{kl}$ being absent. This can happen in the following three cases
\begin{enumerate}
    \item[\hspace{1.5em}- Two winding modes:] $n_i,n_j\neq0$\,,
    \item[\hspace{1.5em}- Three winding modes:] only $n_i=0$, or only $n_j=0$\,,
    \item[\hspace{1.5em}- Four winding modes:] $n_1,n_2,n_3,n_4\neq 0$\,.
\end{enumerate}

In the first case, which will be our guide in what follows, restricting all appearing integers in \eqref{2windings het gen2} to being positive results in an overall prefactor of $2^5\pi$. In the second case, we focus on the first term of \eqref{3windings het general}, which in both cases of $n_i=0$ and $n_j=0$ takes the same form, leading to the total contribution
\begin{equation}
    A^{\rm H}_{\cancel{{E\!F1_{kl}}},2}=2\cdot 2^5\pi\,l_{\rm H}^2V_4^{\rm H}\!\!\!\!\sum_{\substack{N,M,\\m_k,m_l>0\\\hat{n}_i,\hat{n}_j}}\frac{e^{-2\pi N\sqrt{M^2\vartheta_{ij}^2+\hat{n}_i^2m_k^2\vartheta_{ik}^2+\hat{n}_j^2m_k^2\vartheta_{jk}^2+\hat{n}_i^2m_l^2\vartheta_{il}^2+\hat{n}_j^2m_l^2\vartheta_{jl}^2}}}{\sqrt{M^2\vartheta_{ij}^2+\hat{n}_i^2m_k^2\vartheta_{ik}^2+\hat{n}_j^2m_k^2\vartheta_{jk}^2+\hat{n}_i^2m_l^2\vartheta_{il}^2+\hat{n}_j^2m_l^2\vartheta_{jl}^2}}\,,
\end{equation}
where we have appropriately renamed the various integers in order to make comparisons evident. In obtaining this contribution we set one of the momenta in \eqref{3windings het general} to zero and then solve the simple Diophantine equation involving the other two. The winding number that was multiplying this momentum in \eqref{BPSvectorrep} is thus unrestricted and has been renamed to $M$. 

The third case, as expected, is highly non-trivial. In order to bring it into a form appropriate for comparisons to the previous ones, we make the key observation that we may split \eqref{BPSvectorrep} into two terms, namely 
\begin{equation}
    n_im_i+n_jm_j=Q=-n_km_k-n_lm_l\,,
\end{equation}
where $Q=M\cdot {\rm gcd}\left(n_i,n_j\right)\neq0$. The motivation for this manipulation is that, when setting $C_{kl}=0$ in \eqref{vector rep final solution four wind modes}, we observe that 
\begin{equation}
    C_{ij}=M\,,
\end{equation}
which is unconstrained, as we could have expected from \eqref{BPS analogue}. The rest of the  coefficients take precisely the same form as the previous contributions to the same term. This implies that $n_k,n_l,m_i,m_j$ are not directly appearing, but are still constrained due to the BPS condition. 

We perform this computation in detail in appendix \ref{app:5instsolution}, leading to the contribution
    \begin{equation}
    A^{\rm H}_{\cancel{{E\!F1_{kl}}},3}=-2^6\pi\,l_{\rm H}^2V_4^{\rm H}\hspace{-7pt}\sum_{\substack{N,M,\\m_k,m_l>0\\\hat{n_i},\hat{n}_j}}\frac{e^{-2\pi N\sqrt{M^2\vartheta_{ij}^2+\hat{n}_i^2m_k^2\vartheta_{ik}^2+\hat{n}_j^2m_k^2\vartheta_{jk}^2+\hat{n}_i^2m_l^2\vartheta_{il}^2+\hat{n}_j^2m_l^2\vartheta_{jl}^2}}}{\sqrt{M^2\vartheta_{ij}^2+\hat{n}_i^2m_k^2\vartheta_{ik}^2+\hat{n}_j^2m_k^2\vartheta_{jk}^2+\hat{n}_i^2m_l^2\vartheta_{il}^2+\hat{n}_j^2m_l^2\vartheta_{jl}^2}}\,.
\end{equation}

In summary, we can write the total results in terms of identical (but of different physical origin) contributions as
\begin{equation}
    A^{\rm H}_{\cancel{{E\!F1_{kl}}}}=A[n_i,n_j]+A[n_j,n_k,n_l]+A[n_i,n_k,n_l]-2 A[n_1,n_2,n_3,n_4]\,.
\end{equation}
Let us distinguish two cases:
\begin{itemize}
    \item[-] If $i=1$ (analogously $j=1$), then we can interpret our result as the first and third contributions (respectively first and second contributions) canceling that of all the winding numbers turned on.
    \item[-] If $i\neq1\neq j$, then the second and third contributions cancel the contribution from all windings being non-zero and the remaining term does not involve $n_1\neq 0$.
\end{itemize}
 The general term for the bound state of five instantons not wrapping $\vartheta_{kl}$ reads
\begin{equation}
    A^{\rm H}_{\cancel{{E\!F1_{kl}}}}=2\pi\,l_{\rm H}^2V_4^{\rm H}\hspace{-7pt}\sum_{\substack{N>0\\C_{ij},C_{ik},C_{il},C_{jk},C_{jl}\neq 0\\{\rm BPS}}}\frac{e^{-2\pi N\sqrt{C_{ij}^2\vartheta_{ij}^2+C_{ik}^2\vartheta_{ik}^2+C_{il}^2\vartheta_{il}^2+C_{jk}^2\vartheta_{jk}^2+C_{jl}^2\vartheta_{jl}^2}}}{\sqrt{C_{ij}^2\vartheta_{ij}^2+C_{ik}^2\vartheta_{ik}^2+C_{il}^2\vartheta_{il}^2+C_{jk}^2\vartheta_{jk}^2+C_{jl}^2\vartheta_{jl}^2}}\,,
\end{equation}
where again the coefficients satisfy the BPS condition \eqref{BPS analogue}.
As in all previous examples the complete result could be obtained by not including the winding mode $n_1$, which upon translation to type IIA/M-theory variables implies that once again we can confirm the redundancy of $D4$-branes.
 
\subsection{Six heterotic instanton contributions}

The case of a bound state of six instantons can arise in the following cases
\begin{enumerate}
    \item[\hspace{1.5em}- Three winding modes:] any $n_i=0$\,,
    \item[\hspace{1.5em}- Four winding modes:] $n_1,n_2,n_3,n_4\neq 0$\,.
    \end{enumerate}
Given the dual interpretation of $n_1$ winding modes as wrapped $D4$-branes, our emergence arguments would predict that the full result would be obtained by considering only the contribution corresponding to $n_1=0$. Indeed, after reabsorbing the factor of $2^3$ in \eqref{3windings het general} one can verify that this contribution matches with the expected result. 
  To be more precise, inspecting the previous 1-5 instanton cases, one can  express the total worldsheet instanton contribution to the amplitude as
\eq{\label{instsolutiongeneral}
  A^{\rm H}_{\rm inst} = 2\, \pi \,l_{\rm H}^2 V_4^{\rm H}
     \sum_{N>0} \hat{\sum\limits_{C_{ij}|{\rm BPS}}}    \frac{e^{-N S_{\rm inst}(\vec C)}}{S_{\rm inst}(\vec C)}\,,\qquad S_{\rm inst}(\vec{C}):=\sqrt{\sum_{i<j}\vartheta_{ij}^2 \,C_{ij}^2}\,,
}
where the BPS condition is given by \eqref{BPS analogue}. The six instanton contribution predicted by emergence is in perfect agreement with \eqref{instsolutiongeneral}.

However, the precise cancellation is highly involved so we will content ourselves with the observation that the term predicted by emergence is giving rise to the expected result. To be complete, let us summarize the key technical difficulties in formally proving our statement. Although the condition \eqref{BPS analogue} is satisfied and it is trivially obvious in \eqref{3windings het general}, in \eqref{vector rep final solution four wind modes} none of the coefficients are given by simple winding modes. Nevertheless, we observe that three of the coefficients of \eqref{vector rep final solution four wind modes} directly have a form compatible with the BPS condition corresponding to only three winding modes being non-zero, namely
\begin{equation}
    g_3^{-1}\left(n_1C_{23}+n_2C_{13}+n_3C_{12}\right)=0\,,
\end{equation}
so that we could redefine
\begin{align}
    g_3^{-1}C_{23}=M_1&=X_0\mu_1-\hat{n}_2\mu_2\,,\\
 g_3^{-1}C_{13}  = M_2&=Y_0\mu_1+\hat{n}_1\mu_2\,,\\
 g_3^{-1}C_{12}=   M_3&=-\mu_1g_2
\end{align}
and attempt to treat these as the momenta in \eqref{3windings het general}. Then we would manipulate the rest of the coefficients so that the summations over  the winding modes could be exchanged for summations over these coefficients. This turns out to be a daunting task, perhaps signaling already the limitations of our employed techniques.

\section{Wrapped \texorpdfstring{$N\!S5$}{NS5}-branes in five dimensions}
\label{sec4}
So far, we have confirmed one of the main predictions of the M-theoretic emergence proposal, namely the redundancy of heavy longitudinally wrapped $M5$-branes ($D4$-branes) in the case of the $F^4$-coupling. However, another issue which was only tackled on a formal level of group theoretic arguments in \cite{Blumenhagen:2024ydy} was that the light transversally wrapped $M5$-branes ($N\!S5$-branes) are among the fundamental degrees of freedom to be considered.
These become particle-like upon further compactification on a circle.
In this section we provide some comments on the extension of the computation
of  the $F^4$-coupling to such a five-dimensional theory.

Consider again a co-scaled strong coupling limit of type IIA/M-theory compactified on $T^4/\mathbb{Z}_2\times S^1$ such that $M_{\rm Pl}^{(5)}$ does not scale. In this case, transversally wrapped $M5$-branes are particle states that scale as the species scale, which is again the eleven-dimensional Planck scale and following the M-theoretic emergence proposal they should also be integrated out. The scalings of all quantities can be derived by the general formulas of \cite{Blumenhagen:2024ydy,Blumenhagen:2024lmo}. Similarly to the rest of this work, we will take the dual heterotic string theory on a rectangular $T^5$ where the new radius $R_5$ is mapped to type IIA/M-theoretic quantities via
\begin{equation}
\label{radius5}
    R^{\rm H}_5=\frac{1}{g_{\rm IIA}}\sqrt{R_1^{\rm IIA}R_2^{\rm IIA}R_3^{\rm IIA}R_4^{\rm IIA}}R_5^{\rm IIA}=r_{11}\sqrt{v_{T^4}}\,r_5\,,
\end{equation}
while the rest of the quantities are still mapped via \eqref{hetIIradiimap}.

The new states relevant on the heterotic side are the Kaluza-Klein and winding modes along the new direction that, upon mapping to type IIA, give Kaluza-Klein modes along the new direction and $N\!S5$-branes wrapped on $T^4/\mathbb{Z}_2\times S_1$, respectively. Using the duality relations one concludes that all of these states should be integrated out. Identically to the six-dimensional case, the coupling in the heterotic theory is given by the (untwisted) 1/2-BPS partition function $Z_{5,5}$ of 1/2-BPS Kaluza-Klein and winding modes of $T^5$. This can be expressed in terms of the Eisenstein series $\mathcal{E}_{\mathcal{V};\frac{3}{2}}^{SO(5,5)}$ 
subject to the BPS condition $n_i\,m_i=0$, now with $i=1,\ldots,5$. Treating the winding numbers as the coefficients of the Diophantine equation and extending the treatment of appendix \ref{app:Diophantine eqs} for Diophantine equations with five variables we would find a similar momentum contribution encoded in a matrix $\mathcal{M}^{ij}_\alpha$ with det$(\mathcal{M}_\alpha)=L_{{\rm H},\alpha}^2/(V_5^{\rm H})^2$, where $L_{{\rm H},\alpha}^2=\sum_{i=1}^5\tilde{n}_i^2(R_i^{\rm H})^2\alpha_i$, as was proved in \cite{Blumenhagen:2024ydy}.

By repeating the computation of section \ref{secconstantterm},
the constant term obtained from considering the terms involving $n_1\neq 0$, respectively involving $D4$-branes in type IIA, is the same
\begin{equation}
    {\rm C}_{\rm H}^{(5D)}\sim \pi\,l_{\rm H}^2V^{\rm H}_5 \sum_{n_1\neq 0}\sum_{|\alpha|=0}^4(-1)^{|\alpha|}\binom{4}{|\alpha|}\zeta(-1)=0\,,
\end{equation}
though this time we are also considering terms with $n_5\neq0$, thus including $N\!S5$-branes in the dual  type IIA computation. 

A more challenging and non-trivial check of the necessity  of all fundamental degrees of freedom would be to perform an exhaustive study of the worldsheet instantons in five dimensions which  is beyond the scope of the present paper.
However, one can straightforwardly check that emergence considerations are confirmed, for example, in the simplest instanton configuration involving a single instanton. This is possible by direct comparison to six-dimensional results due to the fact that solutions given by more than three non-zero winding numbers only start contributing for bound states of more than one worldsheet instantons. Additionally, using the results of \cite{Blumenhagen:2024ydy}, the worldsheet instanton contributions of $|\alpha|$ non-zero winding numbers in five-dimensions
are appearing with the same initial prefactors as in the six-dimensional case, namely
\begin{equation}\label{ad wind instanton general}
\begin{aligned}
A_{\alpha,{\vec\mu}\neq \vec 0}^{{\rm H}\,(5D)}&\simeq \pi \,l_{\rm H}^2\,V_5^{\rm H}\sum_{\tilde{\mathbf n}_\alpha>0}\sum_{\vec{\mathbf{\mu}}\neq \vec{0}}\sum_{N>0}\frac{2^{|\alpha|}}{L_{{\rm H},\alpha} \sqrt{\mu_i {\cal M}^{-1}_{\alpha,ij}\mu_j}}\,e^{-2\pi N{L_{{\rm H},\alpha} \sqrt{\mu_i {\cal M}^{-1}_{\alpha,ij}\mu_j}}}\,,
\end{aligned}
\end{equation}
where $\mu_i$ are four unconstrained integers and $L_{{\rm H},\alpha}^2=\sum_{i=1}^5\tilde{n}_i^2(R_i^{\rm H})^2\alpha_i$.
An instanton $(E\!F1)_{ij}$ can thus arise by solving the BPS condition with just $n_i,n_j\neq 0$, $n_i\neq 0$ or $n_j\neq 0$. In the cases where $i,j\in\{1,2,3,4\}$ the results follow directly from the previous section and we can formally interpret them as the $n_1\neq 0$ contributions canceling out, thus confirming that again $D4$-branes are mutually canceling.

A new single instanton contribution is that of $(E\!F1)_{15}$ given by solving the BPS condition for $n_1,n_5\neq 0$, $n_1\neq 0$ and $n_5\neq 0$ and again, the contributions with $n_1\neq 0$ cancel each other. On the type IIA side this term is particularly interesting, as after using the relation \eqref{radius5} we obtain the contribution of a Euclidean $E\!D4$-instanton wrapping the entire $T^4/\mathbb{Z}_2\times S^1$.
In this context we note that heterotic worldsheet instantons $(E\!F1)_{i5}$, get mapped to Euclidean $E\!D2$-instantons wrapping appropriate 2-cycles of the $T^4/\mathbb{Z}_2$ together with the fifth direction.
This shows that while particle-like wrapped $D4$-branes are not necessary,
$N\!S5$-branes need to be included in order to get this Euclidean $E\!D4$-instanton contribution.
We expect similar results also for bound states of $E\!D4$-$E\!D2$-$E\!D0$-instantons.

\section{Conclusions}\label{sec5}
In this work we have  provided  detailed evidence for the claim of the M-theoretic emergence proposal that in the  decompactification limit to M-theory only towers of particle-like states
  parametrically not heavier than the species scale are the ones to be integrated out. More specifically, we were focusing on an $F^4$-coupling in six dimensions corroborating that the wrapped $D4$-branes,
featuring  a mass scale larger than  the species scale in the isotropic decompactification limit to M-theory,
are  redundant in the sense of resulting in mutually canceling contributions.
This was achieved by a case-by-case study of the worldsheet instanton contributions to this amplitude on the  dual heterotic side.
We provided the detailed  microscopic cancellation for
bound states of up to five instantons and pointed out some appearing technical difficulties
for the case  of six worldsheet instantons.
Preliminary results in five dimensions also fit within the M-theoretic emergence proposal.

For computational reasons we were working at a very specific
point in the moduli space of a type IIA K3-compactification, where
the internal four-dimensional space  is given by a CFT corresponding to
a $\mathbb Z_2$ toroidal orbifold. This means that there is   half a unit
of Kalb-Ramond flux threading through each of the sixteen blow-up
2-cycles leading to an abelian $U(1)^{16}$ twisted sector gauge group.
It turned out to be very useful that there exists a heterotic dual
description given by a toroidal compactification with a very specific
choice of Wilson lines. It was essential for our work that
at this point in moduli space a couple of cancellations happened \cite{Kiritsis:2000zi}  so that
the 1/2-BPS saturated $F^4$-term for the (type IIA) diagonal twisted sector gauge field
was entirely given by a Schwinger integral with only (heterotic) Kaluza-Klein and winding modes
being integrated out, i.e. the contribution for right-moving string oscillator
modes completely canceled out. This simplification allowed us
to explicitly carry out the detailed evaluation of the appearing
Eisenstein series. For that purpose we could employ
the regularization method of \cite{Blumenhagen:2023tev,Blumenhagen:2024ydy}
with the duality allowing us to freely switch between the heterotic and
the type IIA description.

One should keep in mind that for simplicity our analysis was carried out for rectangular tori where the instantonic actions appearing in the Eisenstein series receive no axion-like contributions.
Moreover, to avoid any confusion let us emphasize that on the heterotic side we were not taking
an emergent string limit but a certain co-scaled decompactification limit
dual to the M-theory limit on the type IIA side. In fact the ten-dimensional heterotic
string coupling was not scaling in this limit at all.
From a technical point of view, the heterotic dual was merely
serving as a powerful  book-keeping device for performing
the actual limit on the type IIA side.

On a conceptual level, it is intriguing that the employed string duality allowed us to utilize
the M-theoretic emergence proposal to reveal a peculiar  microscopic property
of a 1/2-BPS saturated string amplitude, namely the cancellation
of the winding modes along the large direction 1. We consider
this as more evidence for both the M-theoretic emergence proposal
and the general philosophy to perturbatively integrate out
only towers of states with a mass scale not larger than the
species scale. Which specific towers are light and, thus, perturbative in the context of emergence, depends
on the infinite distance limit taken. 
Concerning the M-theoretic emergence proposal,
the final goal is of course to move to more generic, i.e. not supersymmetry protected, amplitudes  
but this requires a complete microscopic description of M-theory
making  this a daunting task momentarily.

\paragraph{Acknowledgments:}
We would like to thank Christian Knei{\ss}l and Joaquin Masias for useful discussions. The work of R.B. is supported  by the Deutsche Forschungsgemeinschaft (DFG, German Research Foundation) under Germany’s Excellence Strategy – EXC-2094 – 390783311.

\appendix
\section{Linear Diophantine Equations}
\label{app:Diophantine eqs}
Diophantine equations are equations with integer coefficients and integer solutions. We will review here the iterative process for solving them that was also used in \cite{Blumenhagen:2024ydy}. More information can be found for example in the classic reference \cite{Mordell}.

For a linear Diophantine equation in two variables, $X$ and $Y$, of the form
\begin{equation}
\label{dioph2var}
a X+ bY = c, \qquad a,b,c \in \mathbb{Z}, \quad \text{$a$ or $b$} \neq 0,
\end{equation}
it is a theorem (e.g.~Theorem 1, chapter 5 of \cite{Mordell}) that integer solutions exist iff gcd$(a,b)$ divides $c$. If $c=0$, the general such solution is $X=Nb$, $Y=-Na$ with $N \in \mathbb{Z}$. If $c\neq 0$, the general solution is 
\begin{equation}
\label{soldioph2}
X = X_0 + Nb\,, \qquad Y = Y_0- Na\,,
\end{equation}
where $(X_0, Y_0)$ is any particular solution of \eqref{dioph2var}.
If $g_2={\rm gcd}(a,b)\neq 1$, one can divide \eqref{dioph2var} by $g_2\equiv{\rm gcd}(a,b)>1$ to get $\tilde a X + \tilde b Y = \tilde c$, with $\tilde a = a/g_2$ and similarly for $\tilde b$, $\tilde c$. Therefore, one recovers again a linear Diophantine equation with coprime coefficients with a solution given by replacing $a \to a/g_2$ and $b \to b/g_2$ in \eqref{soldioph2}. 
One can iterate the algorithm and solve linear Diophantine equations in more variables. 

In this work we mainly focused on linear Diophantine equations in four variables like
\begin{equation}
aX+bY+cZ+dW=0, \qquad a,b,c,d \in\mathbb{Z} \quad \text{with $a$ or $b$ or $c$ or $d\neq 0$}\,.
\end{equation}
From gcd$(a,b,c,d)=1$, we deduce that $W=Pg_3$, with $P\in \mathbb{Z}$.
The general solution is then
\begin{align}
\label{soldioph4v1}
X &=PX_1+MX_0 + N\frac{b}{g_2}\,,\\
Y &=PY_1+MY_0 -N\frac{a}{g_2}\,,\\
Z &=PZ_1 + M g_2\,,\\
\label{soldioph4v2}
W &= Pg_3\,,
\end{align}
with $aX_1+bY_1+cZ_1=-dg_3$,  $g_3= {\rm gcd}(a,b,c)$, and one can check that $aX+bY+cZ+dW=a(X-PX_1)+b(Y-PY_1)+c(Z-PZ_1)=a(MX_0+Nb/g_2)+b(MY_0-Na/g_2)+cMg_2=0$. Observe that to solve an equation in $k$ variables we need to introduce $k-1$ arbitrary integers. 

In \cite{Blumenhagen:2024lmo}, it was observed that when this method is applied to solving the BPS conditions for Kaluza-Klein and winding modes, one can express the Kaluza-Klein contributions to the corresponding Schwinger-like integrals appearing in Eisenstein series as $\sum_{i,j}\mu_i {\cal M}^{ij}_\alpha \mu_j$, where

\begin{equation}
\label{MMatrix5}
{\cal M}^{ij}_\alpha=\resizebox{!}{.07\textwidth}{$\begin{pmatrix}
\frac{\hat{n}_2^2}{R_1^2}+\frac{\hat{n}_1^2}{R_2^2}& \frac{\hat{n}_2X_0}{R_1^2}-\frac{\hat{n}_1Y_0}{R_2^2}&\frac{\hat{n}_2X_1}{R_1^2}-\frac{\hat{n}_1Y_1}{R_2^2}\\
\frac{\hat{n}_2X_0}{R_1^2}-\frac{\hat{n}_1Y_0}{R_2^2}&
\frac{X_0^2}{R_1^2}+\frac{Y_0^2}{R_2^2}+\frac{g_2^2}{R_3^2}&
\frac{X_0X_1}{R_1^2}+\frac{Y_0Y_1}{R_2^2}+\frac{g_2Z_1}{R_3^2}
\\
\frac{\hat{n}_2X_1}{R_1^2}-\frac{\hat{n}_1Y_1}{R_2^2}&
\frac{X_0X_1}{R_1^2}+\frac{Y_0Y_1}{R_2^2}+\frac{g_2Z_1}{R_3^2}&\frac{X_1^2}{R_1^2}+\frac{Y_1^2}{R_2^2}+\frac{Z_1^2}{R_3^2}+\frac{g_3^2}{R_4^2}
\end{pmatrix}$}.
\end{equation}
One sees that ${\rm det}({\cal M}_\alpha)=L_\alpha^2/(V_4)^2$\,, where $L_\alpha^2=\sum_{i=1}^4n_i^2R_i^2\alpha_i$.

\section{Derivation of the five instanton contribution}
\label{app:5instsolution}

In this appendix we provide the details of the derivation of the contribution of the general solution \eqref{vector rep final solution four wind modes} to a bound state of five instantons in our amplitude. In particular, we will work out the case where the term $\vartheta_{12}$ is absent, as any other contribution can be obtained upon changing the order of terms in the BPS condition \eqref{vector rep final solution four wind modes}. The general solution in this case is given by setting $\mu_1=0$ and it can be reexpressed as
\eq{
  \left(L_{\rm H}\sqrt{\mu_i\mathcal{M}_{ij}^{-1}\mu_j}\right)^2&= \vartheta_{13}^2\hat{n}_1^2m_4^2+\vartheta_{23}^2\hat{n}_2^2m_4^2+\vartheta_{14}^2\hat{n}_1^2m_3^2\\ &+\vartheta_{24}^2\hat{n}_2^2m_3^2+\vartheta_{34}^2(\hat{n}_1m_1+\hat{n}_2m_2)^2\,,}
by reintroducing the solutions for the $m$'s obtained in appendix \ref{app:Diophantine eqs} with $N=0$, $M=-\mu_3$ and $P=\mu_2$.

We would like to relate this contribution to that coming from two winding numbers being non-zero, so let us re-express the greatest common divisor of the winding numbers as
\begin{equation}
    g_4={\rm gcd}({\rm gcd}(n_1,n_2),{\rm gcd}(n_3,n_4))\,.
\end{equation}
Because $\hat{n}_3$ and $\hat{n}_4$ do not appear in our general solution we can work with
\begin{equation}
 \sum_{\mathbf{\tilde{n}_4}}\ldots=\sum_{\left(g_2^{(12)},g_2^{(34)}\right)=1}\sum_{\hat{n}_1,\hat{n}_2}\sum_{\hat{n}_3,\hat{n}_4}\ldots\,,
\end{equation} 
where $g_2^{(ij)}={\rm gcd}\left(\frac{n_i}{g_4},\frac{n_j}{g_4}\right)$ so that the hatted integers are pairwise coprime. At this stage, we can take advantage of the fact that these particular greatest common divisors are not appearing in the summand, allowing us to perform the following redefinitions
\begin{equation}
    \sum_{\left(g_2^{(12)},g_2^{(34)}\right)=1}\sum_{\hat{n}_3,\hat{n}_4}\ldots=\frac{1}{\zeta(0)}\sum_{g_2^{(12)}}\sum_{g_2^{(34)}}\sum_{\hat{n}_3,\hat{n}_4}\ldots=-2\sum_{g_2^{(12)}}\sum_{n_3,n_4>0}\ldots\,.
\end{equation}
Effectively, we are using the fact that any unconstrained integer could be the greatest common divisor of the four integers at hand and such a redefinition does not add new factors in our summand. These manipulations lead us to the following contribution
\begin{equation}
    A^{\rm H}_{\cancel{{E\!F1_{12}}}}=-2^5\pi\, l_{\rm H}^2V_4^{\rm H}\hspace{-12pt}\sum_{\substack{N>0\\M,m_3,m_4\neq0\\\hat{n}_1,\hat{n}_2}}\sum_{\substack{g_2^{(12)}>0\\n_3,n_4>0\\{\rm BPS}}}\hspace{-4pt}\frac{e^{-2\pi N\sqrt{M^2\vartheta_{34}^2+\hat{n}_1^2m_3^2\vartheta_{13}^2+\hat{n}_1^2m_4^2\vartheta_{14}^2+\hat{n}_2^2m_3^2\vartheta_{23}^2+\hat{n}_2^2m_4^2\vartheta_{24}^2}}}
    {\sqrt{\ldots}}
    \,,
\end{equation}
where we have introduced the integer $M=n_1m_1+n_2m_2$ for which there is no direct restriction and have avoided copying the square root of the exponent. Note that even though $\hat{n}_3$ and $\hat{n}_4$ do not appear, they are constrained implicitly by the BPS condition $n_1m_1+n_2m_2=-n_3m_3-n_4m_4$. In order to integrate them out we solve
\begin{equation}
    n_3m_3+n_4m_4=-Mg_2^{(12)}=-Q\,,
\end{equation}
for any value of $M,m_3,m_4$. For this to be possible, we demand that $Q=-N{\rm gcd}(m_3,m_4)$. Observe that this demand can always be satisfied without constraining the $M,m_3,m_4$ by 
\begin{equation}
    \left(N,g_2^{(12)}\right)=A(M,{\rm gcd}(m_3,m_4))\,,\quad A\neq0\,.
\end{equation}

We then extend the summations over $g_2^{(12)},n_3,n_4$ to $\mathbb{Z}^*$, restrict $m_3,m_4$ to positives and express $n_3,n_4$ as 
the solutions of \begin{equation}
    n_3\tilde{m}_3+n_4\tilde{m}_4+AM=0\,,
\end{equation}
which take the familiar form 
\begin{equation}
    n_3=\tilde{m}_3P+AM\chi_0\,,\quad
    n_4=-\tilde{m}_4P+AM\psi_0\,,
 \end{equation}
where $\tilde{m}_3\chi_0+\tilde{m}_4\psi_0=-1$ and, in principle, $A\neq0,P\in\mathbb{Z}$. However, in order to be consistent we need to exclude all values of $A,P$ such that any of the two winding numbers is zero. This brings us to yet another Diophantine equation, namely 
\begin{equation}
n_3=0\Rightarrow    (P,A)=B(-M\chi_0,\tilde{m}_3)\,,\quad B\neq0
\end{equation}
and similarly for $n_4=0$. These two sets of solutions do not contain common elements so we can subtract them separately. In fact, their contributions will eventually be equal. Observe that due to the nested structure of these solutions for every value of $B,M,m_l,m_k$ we obtain a unique value of $g_2^{(12)}=Bm_3$ or $g_2^{(12)}=-Bm_4$. Schematically, we obtain
\begin{equation}
    \sum_{\substack{g_2^{(12)},n_3,n_4\neq0}}1=\sum_{A{\rm gcd}(m_3,m_4)}\sum_{\substack{P\in\mathbb{Z}\\A\neq0}}1-2\sum_{B\neq0}=-4\zeta(0)=2\,,
\end{equation}
where we used that $\sum_{P\in\mathbb{Z}}1=0$.
We can thus express this contribution as 
\begin{equation}
    A^{\rm H}_{\cancel{{E\!F1_{12}}}}=-2^6\pi\, l_{\rm H}^2V_4^{\rm H}\hspace{-5pt}\sum_{\substack{N,M,\\m_3,m_4>0\\\hat{n}_1,\hat{n}_2}}\hspace{-3pt}\frac{e^{-2\pi N\sqrt{M^2\vartheta_{34}^2+\hat{n}_1^2m_3^2\vartheta_{13}^2+\hat{n}_1^2m_4^2\vartheta_{14}^2+\hat{n}_2^2m_3^2\vartheta_{23}^2+\hat{n}_2^2m_4^2\vartheta_{24}^2}}}{\sqrt{M^2\vartheta_{34}^2+\hat{n}_1^2m_3^2\vartheta_{13}^2+\hat{n}_1^2m_4^2\vartheta_{14}^2+\hat{n}_2^2m_3^2\vartheta_{23}^2+\hat{n}_2^2m_4^2\vartheta_{24}^2}}\,.
\end{equation}

\newpage

\bibliography{references}  
\bibliographystyle{utphys}

\end{document}